\documentclass[reprint,floatfix,notitlepage,nofootinbib,twocolumn,superscriptaddress,prb]{revtex4-1}
\usepackage{amsmath}
\usepackage{amssymb}
\usepackage{graphicx}
\usepackage[tight]{subfigure}
\usepackage{enumitem}
\usepackage{soul}
\usepackage{cancel}
\usepackage{tikz}
\usepackage{pifont}
\usepackage{xspace}
\usepackage{comment}
\usepackage{algpseudocode}
\usepackage{algorithm}
\usepackage{simplewick}

\makeatletter
\renewcommand{\p@subsection}{}
\renewcommand{\p@subsubsection}{}
\makeatother

\usepackage[english]{babel}
\makeatletter
\def\bbl@set@language#1{%
  \edef\languagename{%
    \ifnum\escapechar=\expandafter`\string#1\@empty
    \else\string#1\@empty\fi}%
  \@ifundefined{babel@language@alias@\languagename}{}{%
    \edef\languagename{\@nameuse{babel@language@alias@\languagename}}%
  }%
  \select@language{\languagename}%
  \expandafter\ifx\csname date\languagename\endcsname\relax\else
    \if@filesw
      \protected@write\@auxout{}{\string\select@language{\languagename}}%
      \bbl@for\bbl@tempa\BabelContentsFiles{%
        \addtocontents{\bbl@tempa}{\xstring\select@language{\languagename}}}%
      \bbl@usehooks{write}{}%
    \fi
  \fi}
\newcommand{\DeclareLanguageAlias}[2]{%
  \global\@namedef{babel@language@alias@#1}{#2}%
}
\makeatother
\DeclareLanguageAlias{en}{english}

\usetikzlibrary{arrows}
\usetikzlibrary{intersections}
\usetikzlibrary{shapes.geometric}
\usetikzlibrary{decorations.pathmorphing, patterns,shapes,fixedpointarithmetic}
\usetikzlibrary{decorations.markings}

\tikzset{
  mid arrow/.style={postaction={decorate,decoration={
        markings,
        mark=at position .575 with {\arrow{stealth}}
      }}},
  end arrow/.style={postaction={decorate,decoration={
        markings,
        mark=at position 1 with {\arrow{stealth}}
      }}},
  snake arrow/.style={fixed point arithmetic, decorate, decoration={snake,amplitude=2pt, segment length=11pt},postaction={decoration={markings,mark=at position 0.625 with {\arrow{stealth}}},decorate}},
}

\newcommand{\red}[0]{\color{red}}
\newcommand{\blue}[0]{\color{blue}}

\newcommand{\I}{\mathbb{I}}
\newcommand{\tr}{\text{tr}}

\usepackage{bm}

\newcommand{\ket}[1]{|#1 \rangle}
\newcommand{\bra}[1]{\langle #1|}
\usepackage[pdfusetitle,colorlinks=true,citecolor=blue,linkcolor=magenta]{hyperref}

\graphicspath{{./}{./images/}}
\begin{document}
\title{Measurement induced entanglement transition in two dimensional shallow circuit}
\author{Hanchen Liu}
\email{liubve@bc.edu}
\affiliation{Department of Physics, Boston College, Chestnut Hill, MA 02467, USA}
\author{Tianci Zhou}
\affiliation{Center for Theoretical Physics, Massachusetts Institute of Technology, Cambridge, Massachusetts 02139, USA}
\author{Xiao Chen}
\email{chenaad@bc.edu}
\affiliation{Department of Physics, Boston College, Chestnut Hill, MA 02467, USA}

\begin{abstract}
We prepare two dimensional states generated by shallow circuits composed of (1) one layer of two-qubit CZ gate or (2) a few layers of two-qubit random Clifford gate. After measuring all of the bulk qubits, we study the entanglement structure of the remaining qubits on the one dimensional boundary. In the first model, we observe that the competition between the bulk X and Z measurements can lead to an entanglement phase transition between an entangled volume law phase and a disentangled area law phase. We numerically evaluate the critical exponents and generalize this idea to other qudit systems with local Hilbert space dimension larger than 2. In the second model, we observe the entanglement transition by varying the density of the two-qubit gate in each layer. We give an interpretation of this transition in terms of random bond Ising model in a similar shallow circuit composed of random Haar gates. 
\end{abstract}

\maketitle

\section{Introduction}
Quantum entanglement is essential to many-body quantum physics and quantum information processing. 
An entangled state can be prepared through a quantum circuit. By applying a series of unitary gates on an array of qubits, we can realize useful quantum states for different computational purposes.
For instance, quantum approximate optimization algorithm (QAOA) makes use of two types of unitaries to construct non-trivial quantum states that produce approximate solutions of combinatorial optimization problems\cite{farhi2014quantum}. Another important class of models is the random circuits with local two-qubit gates, which can efficiently approximate the pseudo-randomness of a Haar circuit in a polynomial depth\cite{Fernando_2012}. These random circuits are important for many sampling tasks in quantum computing and can be used to demonstrate the quantum supremacy \cite{Bouland_2018}. %Sampling problems of these random circuits are tools to benchmark the advantage of quantum computing
Recently, researchers also consider unitary circuit interspersed with non-unitary measurement gates. Such type of hybrid circuit effectively describes a monitored quantum dynamics. It is shown that in this model, there is a generic entanglement phase transition from a highly entangled volume law  phase to a disentangled area law phase by tuning the measurement rate\cite{skinner2019measurement,Li_2018,Chan_2019,gullans2020dynamical,choi2020quantum}.
The non-unitary circuit significantly enriches the family of the dynamically generated quantum phases.
Much progress has been made by using the repeated measurements to protect critical phases or symmetrically/topologically non-trivial phases in quantum dynamics\cite{Chen_2020,alberton2020trajectory,bao2021symmetry,lavasani2020measurementinduced,sang2020measurement,ippoliti2020entanglement,Nahum_Skinner_2020,han2021measurement}.

Alternatively, we can generate various entangled states by merely performing local measurements on the resource states\cite{Raussendorf_2003,Tzu-Chieh_2012}. In this protocol, although the measured qubits are disentangled with the system, the remaining qubits can become more entangled with each other after a measurement. One important example is the measurement based quantum computing, in which the computation is realized by performing local measurement on an initially prepared resource state\cite{Raussendorf_2003,Tzu-Chieh_2012}. 
%It is shown that any unitary quantum circuit can be mapped to a measurement pattern on the qubits of the resource state\cite{Raussendorf_2003}. 
A massive overhead is required in this approach, since all of the measured qubits are discarded after the computation. Another example is the tensor network, which is a efficient numerical tool for representing non-trivial many-body states and simulating quantum circuits\cite{Orus_2014,Verstraete_2006,perezgarcia2007matrix}. In this approach, an entangled quantum state is constructed by contracting elementary tensors. Such contractions can be effectively treated as local measurements.

There has been a growing interest in the tensor networks in the past few years. 
Among all these developments, the random tensor network has drawn much attention due to its application in quantum gravity and quantum information\cite{Jahn_2021,hayden2016holographic}. For example, consider a two dimensional (2d) tensor network in which each random tensor is a gaussian random states.  By contracting these tensors in the 2d bulk, a one dimensional boundary state can be generated\cite{hayden2016holographic}. When the bond dimension $q$ of each tensor goes to infinity, the entanglement entropy of the boundary state saturates to the minimal cut formula and provides a nice geometric demonstration of the holographic duality\cite{hayden2016holographic,Ryu_Takayanagi}. Decreasing the bond dimension suppresses the entanglement and when $q<q_c$, the boundary state is in the disentangled area law phase\cite{Vasseur2019}. Recently, this problem has been revisited in the random stabilizer tensor network defined on the rectangular lattice, where large-scale numerical simulation confirms the existence of an entanglement transition by tuning $q$\cite{yang2021entanglement}. Besides this transition, it is discovered that when this tensor network is further subject to single qudit bulk measurement, there exists a {\it continuous} entanglement phase transition by varying the measurement rate, akin to the measurement induced phase transition in the hybrid circuit\cite{yang2021entanglement}. 

The discovery of this continuous transition motivates us to ask the following question: For a 2d wave function, after performing measurement for the bulk degrees of freedom, can the remaining 1d boundary state have interesting entanglement structure? In particular, by tuning some parameters in the bulk, can the boundary state undergo an entanglement phase transition? In this paper, we will show that the answer is yes in a few circuit models. Instead of constructing 2d tensor network, we directly prepare a 2d state generated by a {\it shallow} circuit. Although this 2d state is area law entangled, the measurement in the bulk can potentially induce an entanglement transition on the boundary. We consider shallow circuit composed of one layer of controlled phase gates. For every qudit in the bulk, we perform $X$ measurement with probability $p_x$ and $Z$ measurement with probability $1-p_x$. Since $X$ measurement tends to entangle the neighboring qudits and $Z$ measurement disentangles its neighbors, increasing $p_x$ can induce an entanglement phase transition from the area law entangled state to the volume law entangled state for the boundary state. We numerically compute this transition by using the stabilizer formalism and extract critical exponents around the critical points. We further consider a shallow circuit composed of a few layers of random two-qubit Clifford gates. We find that by varying the density of two-qubit gates in each layer, there also exists an entanglement phase transition for the post-measurement boundary state. To understand this phase transition, we consider a similar shallow circuit composed of random Haar gates. We argue that in the replicated Hilbert space, this transition (at least at large local Hilbert space dimension limit) can be mapped to an order-disorder phase transition in the random bond Ising model.
In the above analysis, the circuit depth needs to be shallow, otherwise the measurement in the bulk qudits may not be able to affect the scaling of the entanglement of the boundary qudits. From the computational complexity perspective, a similar entanglement phase transition in the random shallow circuit has been studied in Refs.~\onlinecite{napp2020efficient,bao2022finite}.

The rest of this paper is organized as follows: In Sec.~\ref{sec:graph}, we first consider the graph state generated by one layer of CZ gates. We study the measurement induced entanglement phase transition in the 1d boundary state and then generalize this idea to qudit case. In Sec.~\ref{sec: random_clifford}, we study the boundary entanglement phase transition in shallow circuit composed of random Clifford gate. We give an interpretation of this transition by considering a similar random Haar circuit in Sec.~\ref{sec:haar}. We conclude the main results and discuss possible future research direction in Sec.~\ref{sec:conclusion}.

%In the graph state, the $Z$ measurement breaks the edges while $X$ can create new edges. The competition between them can lead to an entanglement phase transition.

%We focus on the Clifford circuit since we can easily simulate the highly entangled volume law phase on the classical computer.

% The general idea of our method goes as follows.
 
%This approach has potential application on the observation of entanglement phase transition. However this requires massive overhead/ has high overhead. 

\section{Measurement induced entanglement transition in the graph state}
\label{sec:graph}
%In this section, we consider a two dimensional many-qubit stabilizer state $|\psi\rangle$ generated by a 2d shallow circuit composed of local two-qubit unitary Clifford gates. We first investigate a circuit composed of solely CZ gates and then generalize it to the random Clifford gates. In both cases, we perform single qubit Pauli measurements for the bulk qubits and study the possible entanglement phase transition for the 1d boundary qubits.

In this section, we consider 2d graph state generated by one layer of CZ gates. We first consider qubit systems and then generalize to qudit systems. In both cases, we perform single qubit/qudit measurements for the bulk of the graph state and study the possible entanglement phase transition for the 1d boundary qubits/qudits.

\subsection{Qubit graph state}
\subsubsection{Review of stabilizer formalism}

An N-qubit stabilizer state can be defined as the simultaneous eigenstate of $N$ commuting and independent Pauli string operators with eigenvalue +1. These Pauli strings form the generators of the stabilizer group and completely define the wave function $|\psi\rangle$. Since each Pauli string $P_n$ can be written as $\prod_{i=1}^N X^{a^n_i}_iZ^{b^n_i}_i$ with $a_i,b_i=$ $0$ or $1$, the information of the entire wave function can be conveniently stored in a $N\times 2N$ binary matrix $T=[T_X, T_Z]$, where $T_X$ and $T_Z$ are both square $N\times N$ matrices. In this stabilizer tableau $T$, the $n$th row of $T_X$ and $T_Z$ encode the information of $\{a^n_i\}$ and $\{b^n_i\}$ respectively. For a stabilizer state evolved under the Clifford gates, its stabilizer generators will be transformed into a new set of Pauli strings with the $T$ matrix updated accordingly. The stabilizer formalism provides a very efficient method for simulating Clifford dynamics and analyzing properties of the stabilizer state on the classical computer\cite{gottesman1998heisenberg,Aaronson_2004}. In particular, for the stabilizer state, the R\'enyi entanglement entropy $S_A=\frac{1}{1-n}\log_2\mbox{Tr}\rho_A^n$ of the subsystem A obeys the form\cite{Hamma}
\begin{align}
    S_A=\mbox{rank}_2 (T_A)-N_A
\end{align}
where $\mbox{rank}_2 (T_A)$ is the binary rank for the truncated stabilizer tableau $T_A$ in the subsystem A. Notice that in the stabilizer state, $S_A$ is independent of the R\'enyi index $n$.

In the stabilizer state, there is a subset of wave functions in which $T_X$ is an identity matrix\cite{Raussendorf_2003}. Since all of the stabilizers commute with each other, $T_Z$ is required to be a binary symmetric square matrix, which is also an adjacency matrix for a undirected graph. For this reason, this subset of wave function is denoted as the graph state. The graph state can be generated by first preparing a state with all the qubits polarized in the $x$ direction, i.e., 
\begin{align}
|\psi_0\rangle=|+\rangle^{\otimes |V|}
\label{eq:polarized_x}
\end{align}
where the qubits are living on the vertices $V$ of the graph $G$. In $|\psi_0\rangle$, the stabilizer tableau has $T_X$ as an identity matrix and $T_Z$ as a zero matrix. We then apply two-qubit Controlled-Z (CZ) gate (defined in the computational $Z$ basis)
\begin{align}
    {\mbox CZ}=\begin{pmatrix}
    1 & 0 & 0 & 0\\
    0 & 1 & 0 & 0\\
    0 & 0 & 1 & 0\\
    0 & 0 & 0 & -1\\
    \end{pmatrix}
\end{align}
along the edges $E$ of the graph $G$ to construct the graph state. For the CZ gate applied on two qubits pair $(n,m) \in E$, we have $P_m=X_m\to X_mZ_n$ and $P_n=X_n\to X_nZ_m$. As a consequence, the stabilizer generators of such graph state are given by 
\begin{align}
    P_n=X_n\prod_{m|(n,m)\in E} Z_m.
\end{align}
The $T_Z$ matrix is the adjacency matrix for the graph $G=\{V,E\}$ with $T_Z^{nm}=1$ if $(n,m)\in E$ and $T_Z^{nm}=0$ if $(n,m)\notin E$. For a graph state bipartitioning into A and $\overline{\mbox{A}}$ with
\begin{align}
    T_Z=\begin{pmatrix}
    T_Z^{AA} & T_Z^{A\overline{A}}\\
    T_Z^{\overline{A}A} & T_Z^{\overline{AA}} \end{pmatrix},
\end{align}
 the entanglement entropy for subsystem A is\cite{hein2006entanglement}
\begin{align}
    S_A=\mbox{rank}_2(T_Z^{A\overline{A}}),
\end{align}
which also characterizes the connectivity between A and $\overline{\mbox{A}}$ in the corresponding graph.

\subsubsection{Entanglement transition}
We take an initial stabilizer state defined in Eq.\eqref{eq:polarized_x} on the rectangular lattice with the qubits living on the vertices. We then apply a layer of CZ gates along all of the edges between neighboring vertices. Since CZ gates are diagonal in the computational $Z$ basis and commute with each other, we can apply all of them instantaneously on $|\psi_0\rangle$ and  generate a rectangular graph state.  We perform project measurement for the bulk qubits in the $Z$ or $X$ directions and decouple them from the system. The rest of the qubits on the boundary is still a stabilizer state. 

\begin{figure}[h]
\centering
\includegraphics[width=0.45\textwidth]{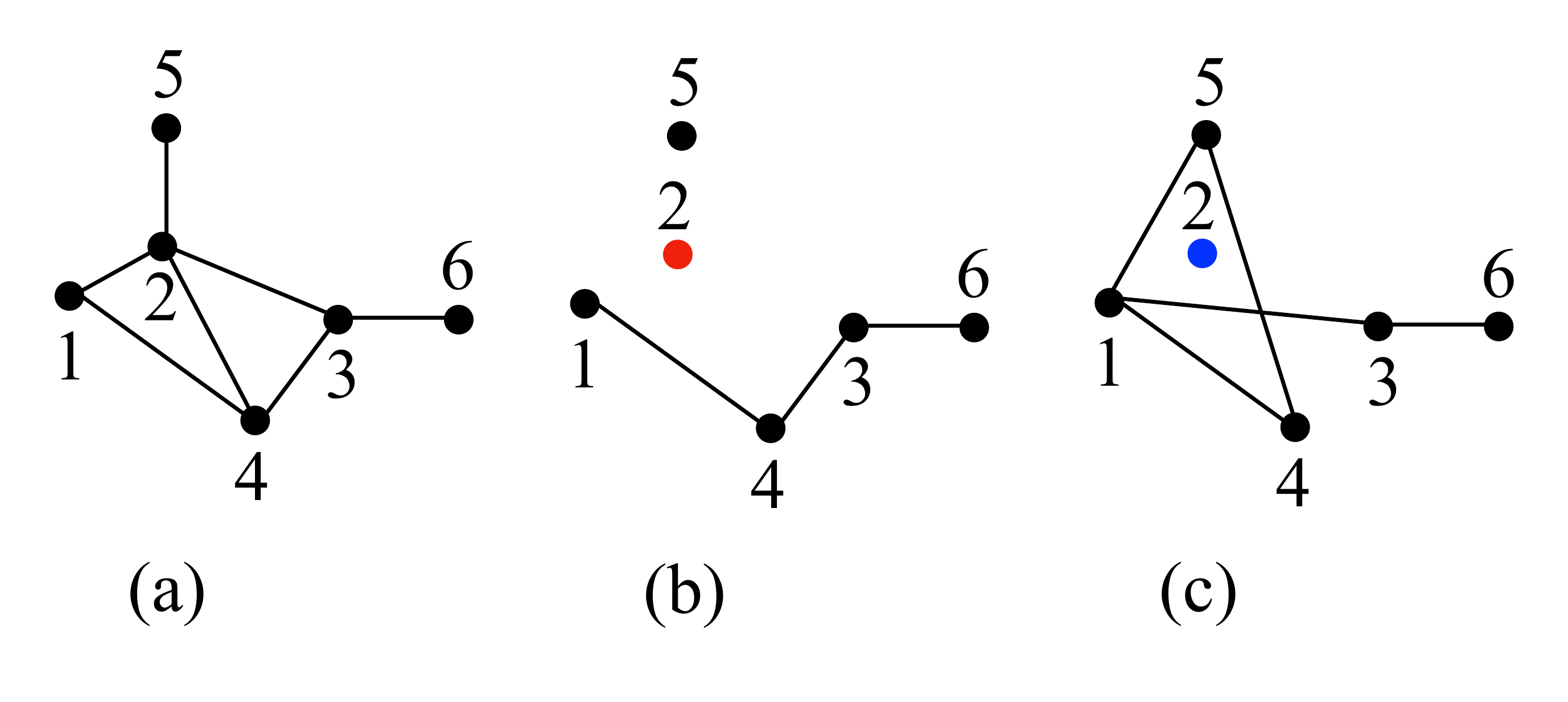}
\caption{Three examples of 6 qubits graph state, where the CZ gates are applied along these bonds. Performing $Z$ measurement on the $2$nd qubit in (a) removes the bonds connecting this qubit with the rest of qubits and leads to the graph state in (b). (c) is the resulting graph state of performing $X$ measurement on the $2$nd qubit followed by applying single qubit Hadamard (H) rotation on the $1$st qubit. Compared with the graph in (b), we have new bonds between $(1,3)$, $(1,5)$ and $(4,5)$ while the bond between $(3,4)$ disappears. }
\label{fig:graph_schematic}
\end{figure}

Before exploring the entanglement scaling of the post-measurement stabilizer state, we analyze the effect of the Pauli $X/Z$ measurements by considering an instance of a six-qubit graph state described in Fig.~\ref{fig:graph_schematic} (a). For a graph state, the Pauli $Z$ measurement at $i$th site simply decouple this qubit by removing the edges connecting this site. As shown in   Fig.~\ref{fig:graph_schematic}(b), the $Z$ measurement at the 2nd site removes the edges between this site and $1$st, $3$rd, $4$th and $5$th sites. In contrast, the $X$ measurement can generate new edges between its neighbors and therefore potentially induce the entanglement between them. In the example described in Fig.~\ref{fig:graph_schematic}(c), after $X$ measurement, the nonzero mutual information is induced between pairs $(3,6)$ and $(4,5)$.

\begin{figure}[h]
\centering
\includegraphics[width=0.45\textwidth]{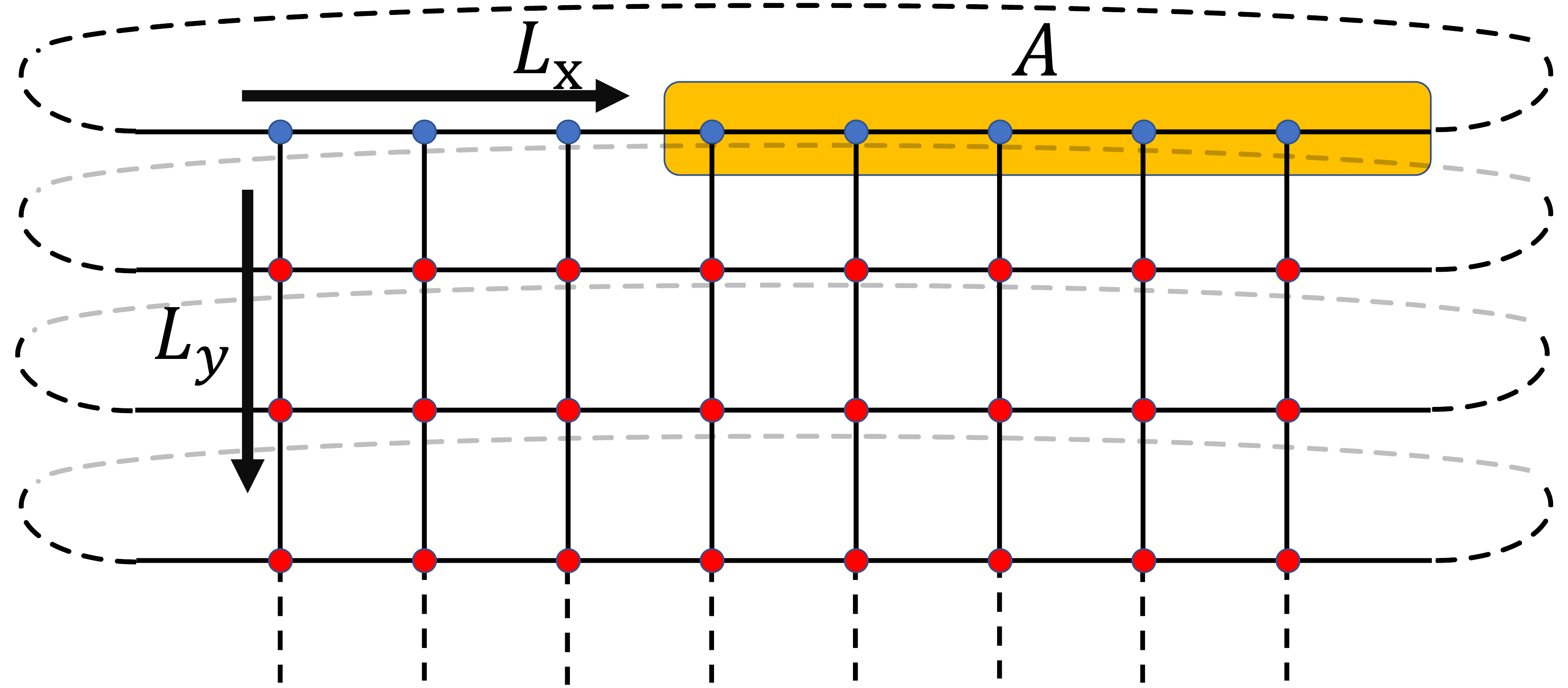}
\caption{The schematics for the $L_x\times L_y$ rectangular lattice  with periodic boundary condition along the $x$ direction. The qubits are living on the vertices of the lattice. The red dots are for the measured qubits and the blue dots are for the un-measured qubits. The orange shaded area denotes the subsystem A at the top boundary. }
\label{fig:cir}
\end{figure}

\begin{figure}
\centering
\includegraphics[width=0.45\textwidth]{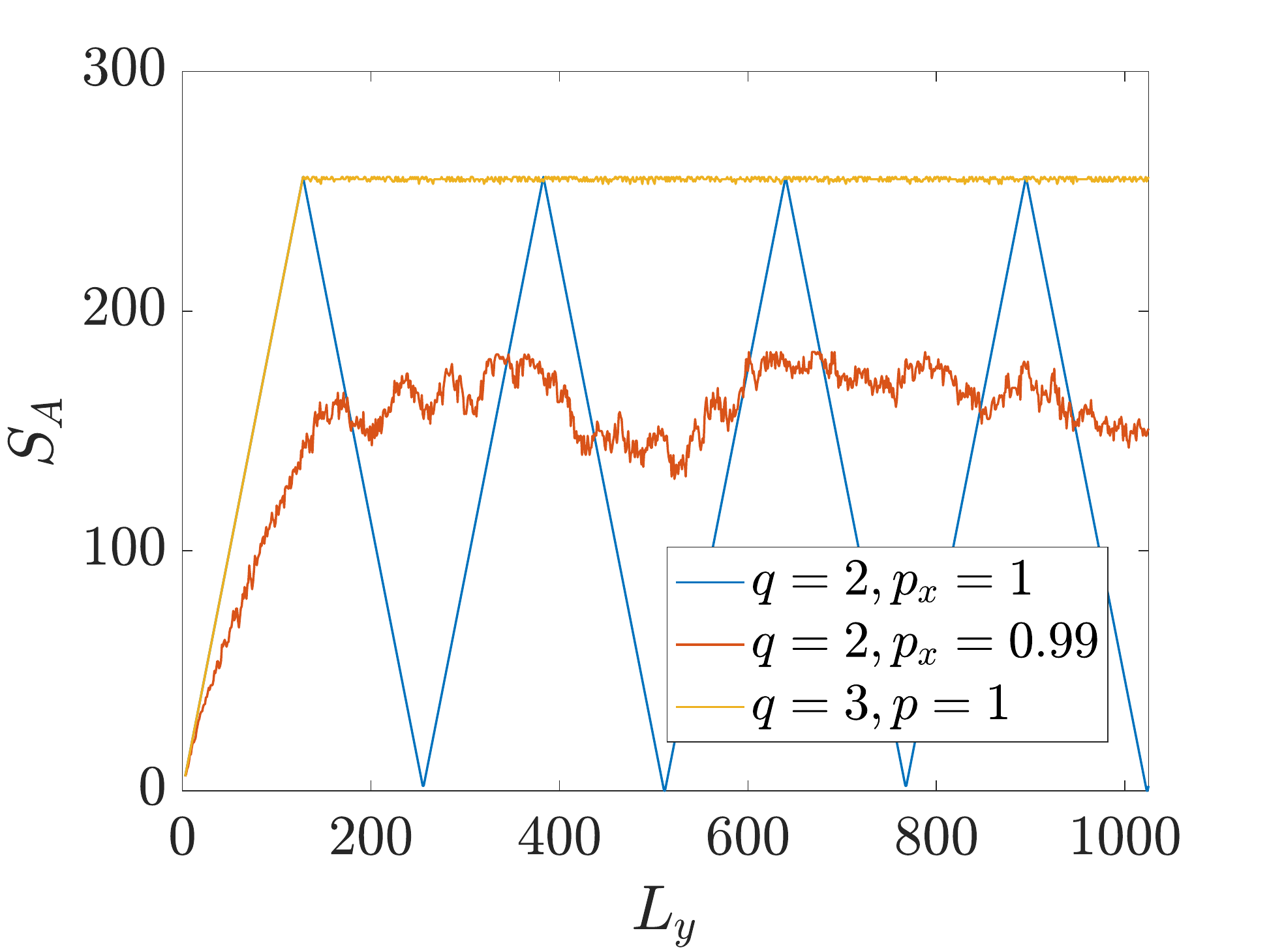}
\caption{ The entanglement entropy $S_A$ as a function of $L_y$ in one circuit realization. The setup is described in Fig.~\ref{fig:cir} with the length of the subsystem $L_A=L_x/2=256$. The periodic oscillation behavior with $p_x=1$ and $q=2$ disappears when $p_x$ is slightly smaller than $1$ or $q>2$.}
\label{fig:EE_time}
\end{figure}

With the rules established in the above illustrative example, we now consider the many-qubit graph state defined on the $L_x\times L_y$ rectangular lattice with periodic boundary condition along the $x$ direction (See Fig.~\ref{fig:cir}). We measure all of the qubits except the ones in the top boundary. For each of these measured sites, we randomly choose $X$ measurement with probability $p_x$ and $Z$ measurement with probability $1-p_x$. In the limit $p_x=0$, the $Z$ measurement simply removes all the edges connecting the measured qubits and the top boundary is a 1d graph state where each site is only connected with two neighboring sites. Such state has $S_A=2$ for a single interval subsystem. On the other hand, when $p_x\to 1$, the $X$ measurement in the bulk qubits can generate edges between its neighbors as demonstrated in Fig.~\ref{fig:graph_schematic} (c). As we increase $L_y$, both the number and length of these edges grow which further leads to the growth of the entanglement entropy in the top boundary. As shown in Fig.~\ref{fig:EE_time}, the entanglement entropy grows linearly in $L_y$ when $p_x$ is very close to 1. The periodic oscillation behavior observed at $p_x=1$ is a special property of the qubit graph state with only $X$ measurement. This behavior disappears for the ``random qudit graph state" with $q\geq 3$, where $S_A$ saturates to a maximally entangled  state when $L_y=L_x/4$ and remains highly entangled when we further increase $L_y$ (The graph state with $q\geq 3$ will be discussed in Sec.~\ref{sec:qudit}). For the qubit graph state, if we take $p_x$ slightly smaller than $1$, the periodic oscillation of $S_A$ is also gone. The fluctuation with $p_x=0.99$ observed in Fig.~\ref{fig:EE_time} is smeared out when we consider sample average. 

The above analysis indicates that $L_y$ could be treated as an effective ``time" direction and the entire 2d graph state subject to Pauli measurements is similar to a 1+1d hybrid circuit dynamics in which the unitary dynamics is interspersed by the local measurement. To efficiently perform the simulation, we do map this 2d problem to a 1+1d dynamics problem. This idea is inspired by the Ref.~\onlinecite{napp2020efficient} where the authors developed a generic algorithm for the output sampling in the shallow circuit with the circuit depth smaller than a critical value. In the stabilizer state, such algorithm allows us to perform large scale simulation, especially with very large $L_y$. The detail of this algorithm is explained in the App.~\ref{sec:algorithm}.

\begin{figure}
\centering
\includegraphics[width=0.45\textwidth]{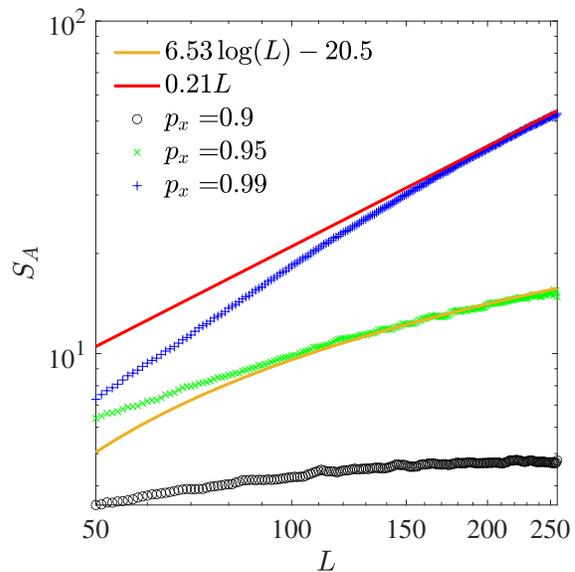}
\caption{Steady state entanglement entropy $S_A$ for various $p_x$ on the log-log scale. Here we take $L_x = L_y = L$ and fix the ratio $L_A/L = 1/4$ and compute $S_A$ as a function of $L$.}
\label{fig:q2EE}
\end{figure}

In the graph state, the $Z$ measurement breaks the edges while $X$ can create new edges. The competition between the bulk $X$ and $Z$ measurements can lead to an entanglement phase transition for the 1d boundary state. We compute the ``steady state" entanglement entropy for the top boundary shown in Fig.~\ref{fig:cir} with large $L_y$. Numerically, we take $L_y\geq L_x$ which is large enough for $S_A$ of the top boundary to  saturate. We observe that there exists an entanglement phase transition by tuning $p_x$. When $p_x<p_x^c=0.95$, $S_A$ is a finite constant independent of the subsystem size. On the other hand, when $p_x^c<p_x<1$, $S_A$ satisfies volume law scaling. The numerical results are presented in Fig. \ref{fig:q2EE}, where we fix the ratio between the subsystem length $L_A$ and $L_x$ to be $1/4$ and plot $S_A$ as a function of $L_x$. 

\begin{figure}
\centering
\includegraphics[width=0.45\textwidth]{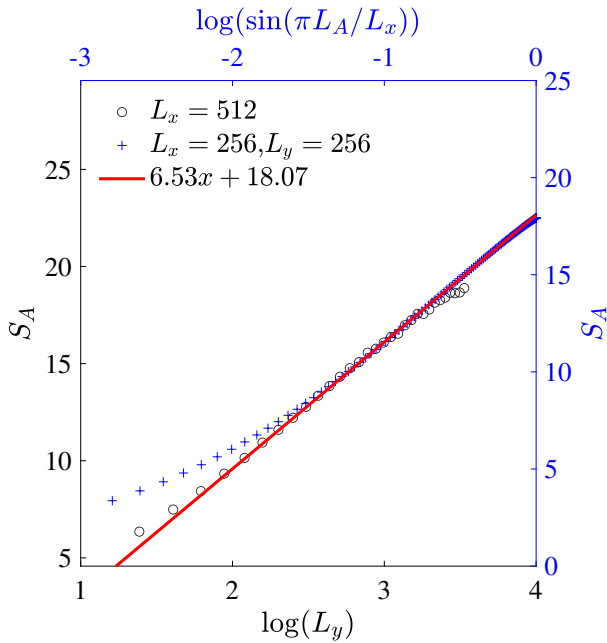}
\caption{Entanglement entropy scaling at critical point $p_x^c$. Blue dots: $S_A$ vs $\log(\sin(\pi L_A/L_x))$. The system size is fixed with $L_x = L_y = 256$. This value of $L_y$ is large enough for the top boundary to saturate to the steady state. Black dots: $S_A$ grows linearly in $\log L_y$ when $L_y\ll L_x$. Here we take $L_A= L_x/2 = 256$. }
\label{fig:q2EEcritLD}
\end{figure}

At the critical point $p_x^c=0.95$, we find that $S_A$ has a logarithmic scaling with the subsystem size $L_A$, the same as that for the steady state of the 1+1d hybrid circuit at the critical point \cite{Li_2019,skinner2019measurement,li2020conformal,Iaconis_2020}. For the periodic boundary condition, we have 
\begin{align}
    S_A=2\alpha\log\left[\frac{L_x}{\pi}\sin(\frac{\pi L_A}{L_x})\right]
    \label{eq:EE}
\end{align}
where $\alpha=3.27$ (See Fig. \ref{fig:q2EEcritLD}). This result is also consistent with Fig. \ref{fig:q2EE}, where we observe that $S_A=2\alpha \log L_x+\cdots$ when the ratio $L_A/L_x$ is fixed. 

\begin{figure}
\centering
\includegraphics[width=0.45\textwidth]{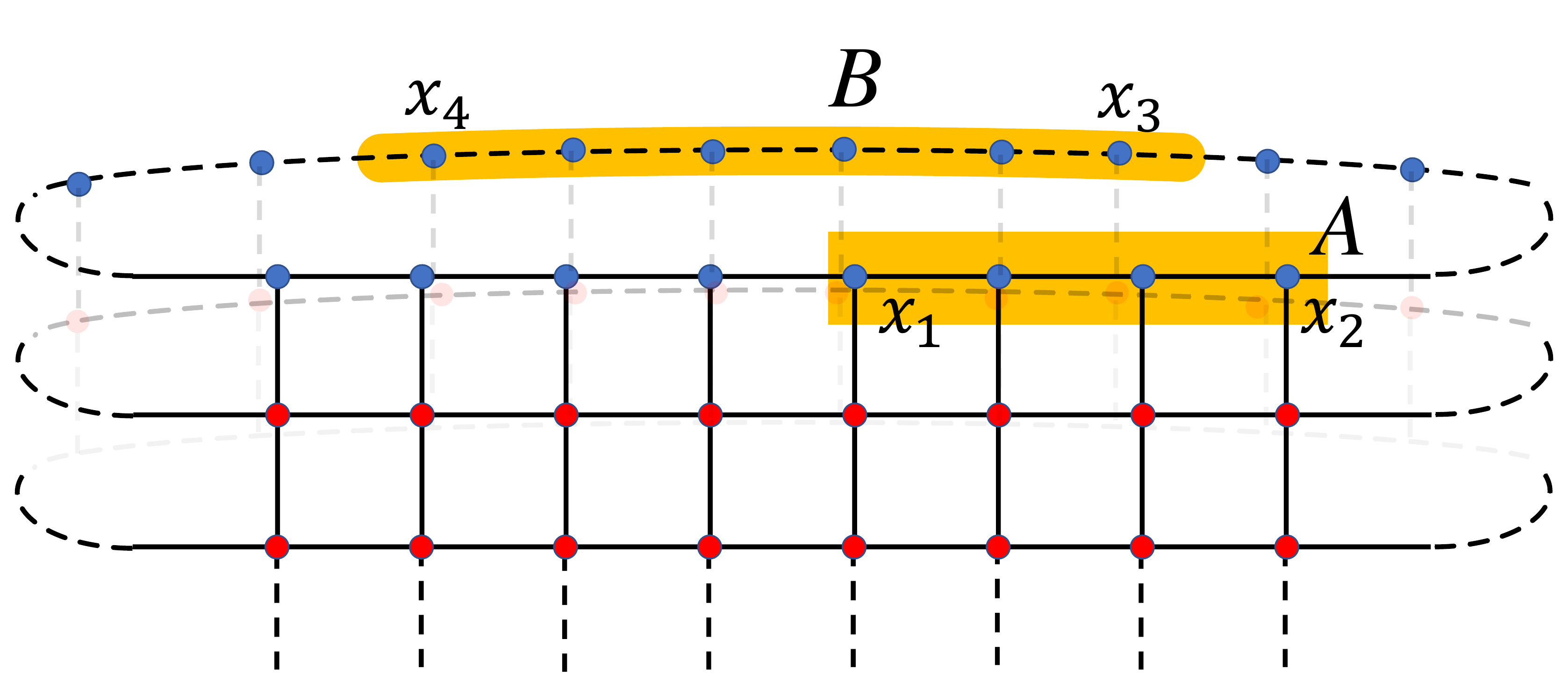}
\caption{The schematics for two randomly chosen area $A = [x_1, x_2]$  and $B = [x_3, x_4]$ (orange shaded areas) in the top boundary.}
\label{fig:mtA}
\end{figure}

\begin{figure}
\centering
\includegraphics[width=0.45\textwidth]{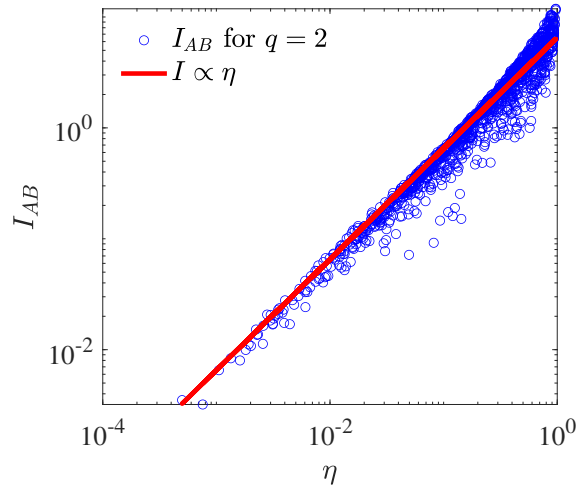}
\caption{The mutual information $I_{AB}$ of two randomly chosen disjoint areas $A = [x_1,x_2]$, $B = [x_3,x_4]$ as a function of cross ratio $\eta$ at critical point $p^c_x = 0.95$ with system size $L_x = L_y = 256$.}
\label{fig:miq2}
\end{figure}

Besides this result, below we also analyze a few other quantities developed in the 1+1d hybrid circuit in Ref.~\onlinecite{Li_2019,li2020conformal,Iaconis_2020} to characterize the critical behaviors at $p_x^c$. We first compute the mutual information between two subsystems A and B defined as 
\begin{align}
    I_{AB}=S_A+S_B-S_{AB}.
\end{align}
Here A and B are two disjoint intervals with $A = [x_1,x_2]$ and $B = [x_3,x_4]$ shown in Fig.~\ref{fig:mtA}. The locations of these four points are randomly chosen on the top boundary. At $p_x^c$, we observe that for the steady state, $I_{AB}$ is a function of cross ratio and does not explicitly depend on $x_i$. This result is presented in Fig.~\ref{fig:miq2}, where all the data points fall on a single curve. Here the cross ratio $\eta$ is defined as
\begin{align}
\eta = \frac{x_{12}x_{34}}{x_{13}x_{24}},\quad \mbox{with}\ x_{ij} = \frac{L_x}{\pi}\sin\left(\frac{\pi}{L_x}|x_i - x_j |\right).
\label{eq:eta}
\end{align}
In particular, we observe that when $\eta\ll 1$, $I_{AB} \sim \eta^\Delta$ with $\Delta\approx 1$. This result indicates that for two small distant regimes, $I_{AB}\sim 1/r^2$, where $r$ is the separation between these two regimes. 

\begin{comment}
\begin{figure}
\centering
\includegraphics[width=0.45\textwidth]{q2DC.eps}
\caption{Data Collapse of steady state $S_A$ around $p_x^c = 0.95$, $\nu = 0.75$}
\label{fig:dcq2}
\end{figure}

We also perform data collapse of the steady state $S_A$ around $p_x^c$ as shown in {\red Fig.\ref{fig:dcq2}}. We fix the ratio $L_A/L$ and find that it satisfies 
\begin{align}
    |S_A(p_x)-S_A(p_x^c)|=f(L|p_x-p_x^c|^\nu)
\end{align}
where {\red $\nu= 0.75$}. {\blue should here be absolute value ? } {\red Yes, we can take absolute value.} This scaling form is the same as that for the 1+1d hybrid circuit except that $\nu$ is much smaller than $\nu$ in the 1+1d hybrid circuit and violates the Harris criterion. 
\end{comment}

In addition, we study $S_A$ as a function of $L_y$ when $L_y \ll L_x$ and we observe that $S_A$ grows logarithmically in $L_y$ with the coefficient as $2\alpha$ (See Fig.\ref{fig:q2EEcritLD}). This behavior is also observed at the critical point of the 1+1d hybrid circuit with 2d conformal symmetry \cite{li2020conformal}.  

\begin{figure}
\centering
\includegraphics[width=0.45\textwidth]{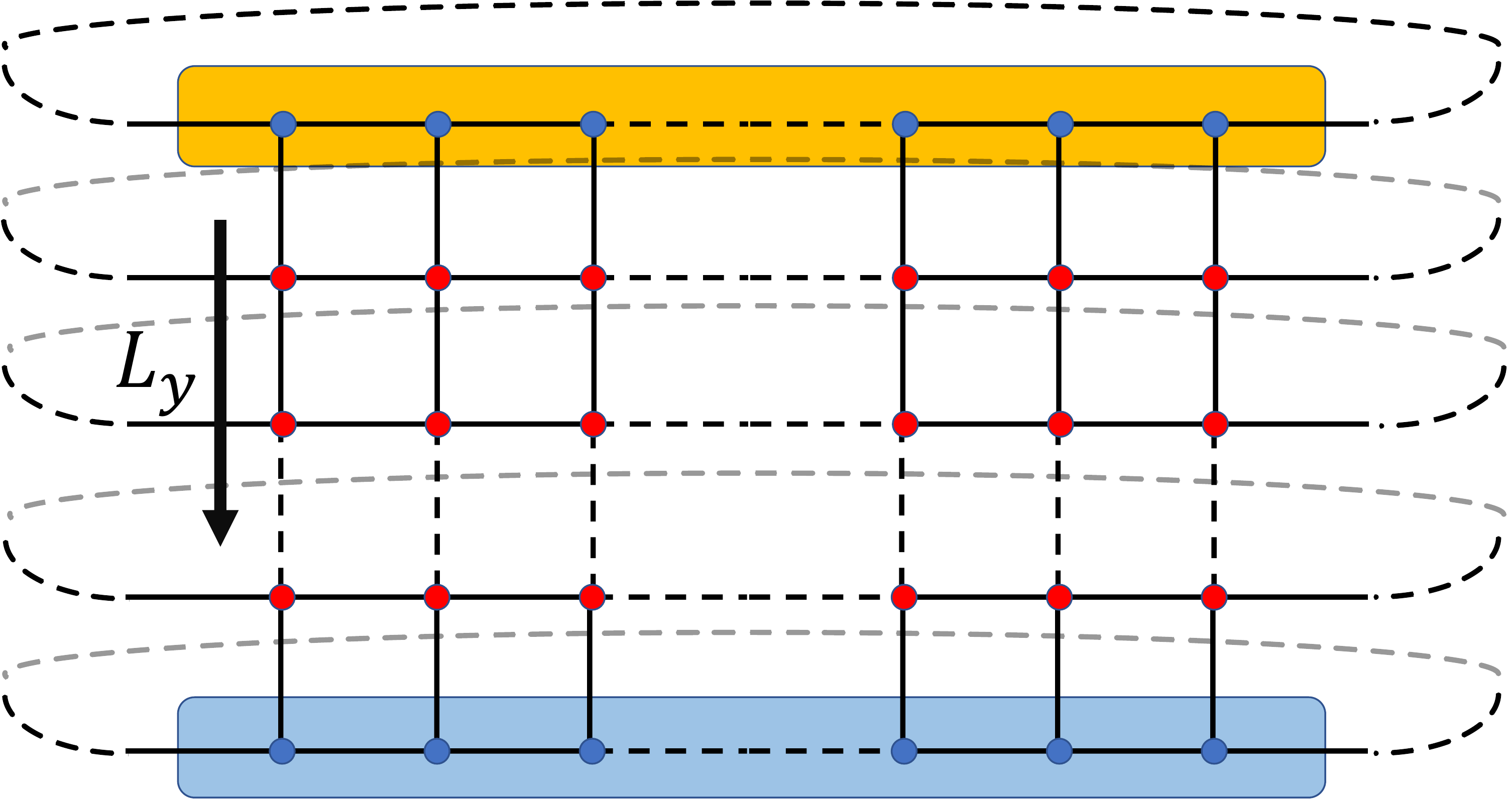}
\caption{The schematics for the cylinder in which both top and bottom qubits are un-measured (blue dots). In this setup, we are interested in the entanglement entropy between the top boundary (orange shaded area) and bottom boundary (blue shaded area) as a function of $L_y$.}
\label{fig:puriDq2}
\end{figure}

\begin{figure}
\centering
\includegraphics[width=0.45\textwidth]{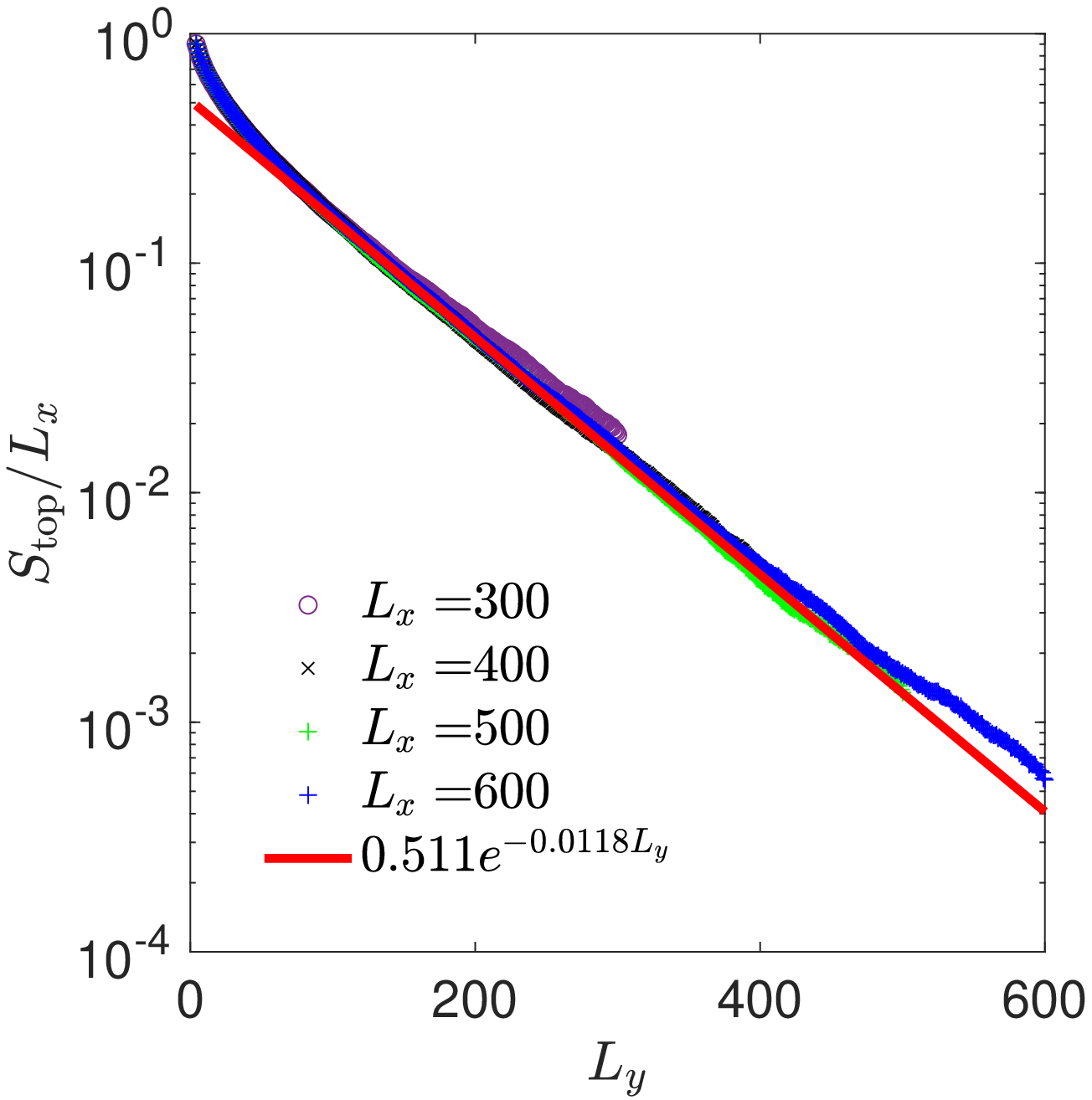}
\caption{
The entanglement entropy $S_{\mbox{top}}$ between the top and bottom boundaries at $p_x^c = 0.95$. $S_{\mbox{top}}$ decays exponentially in $L_y$ for various $L_x$ with the same decay rate.
%The purification dynamics of the qubit system at criticality $p_x = p_x^c = 0.95$. The entanglement entropy between the top and bottom surface $S_{top}$ decays exponentially with $L_y$.
}
\label{fig:puriDq2cData}
\end{figure}

\begin{figure}
\centering
\includegraphics[width=0.45\textwidth]{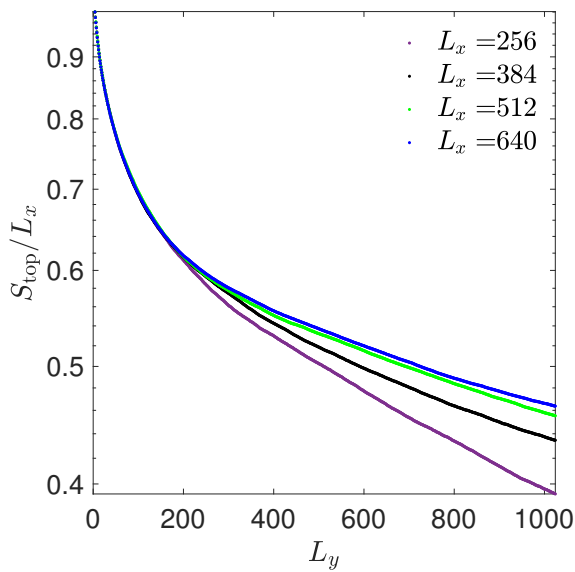}
\caption{
The entanglement entropy $S_{\mbox{top}}$ between the top and bottom boundaries at $p_x = 0.99>p_x^c$. $S_{\mbox{top}}$ decays exponentially with $L_y$ for various $L_x$. The decay rate decreases as we increase  $L_x$.
%The purification dynamics of the qubit system at $p_x = 0.99$ with system size $L_x = 256, 384, 512, 640$. The entanglement entropy between the top and bottom surface decays exponentially with $L_y$. 
}
\label{fig:puriDq2VolData}
\end{figure}

\begin{figure}
\centering
\includegraphics[width=0.45\textwidth]{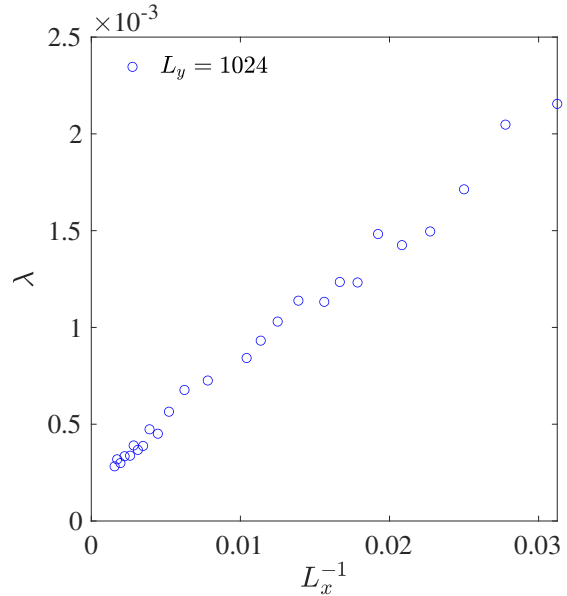}
\caption{Decay rate $\lambda$ (defined in Eq.\eqref{eq:decay_rate}) in the volume law phase as a function of  $L_x^{-1}$. Here we take $p_x = 0.99$ and $ 32 \le L_x \le 640 $ with the maximal $L_y^{\text{max}} = 1024$.}
\label{fig:puriDq2VolDRData}
\end{figure}

Finally, we consider the geometry described in Fig. \ref{fig:puriDq2}, where there are qubits at both the top and bottom boundaries  left un-measured. We compute the entanglement between the top and bottom boundaries $S_{\mbox{top}}$ as a function of $L_y$ and we observe that it decays exponentially fast in Fig. \ref{fig:puriDq2cData}. In particular, the decay rate is a finite constant independent of $L_x$, indicating that the top boundary takes $\log L_x$ time to purify. Such fast decay behavior is distinct from the purification dynamics observed in 1+1d hybrid circuit, where $S_{\mbox{top}}$ is a scaling function $g(L_x/L_y)$ at the critical point\cite{gullans2020dynamical,li2020conformal}. Furthermore, we also study $S_{\mbox{top}}$ in the volume law phase with $p_x>p_x^c$ and we observe that it also decays exponentially in $L_y$, i.e., 
\begin{equation}
    S_{\mbox{top}}/L_x \propto \exp(-\lambda L_y)
    \label{eq:decay_rate}
\end{equation}
as shown in Fig.~\ref{fig:puriDq2VolData}.
Different from $p_x=p_x^c$, the decay rate $\lambda$ is a function of $L_x$ and decreases as we increase $L_x$. As shown in Fig.~\ref{fig:puriDq2VolDRData}, the numerical simulation for $32\le L_x \le 640$ indicates that the decay rate $\lambda$ depends linearly on $L_x^{-1}$. Such fast decay behavior in both volume law phase and critical point implies that the entanglement phase transition in the graph state may not be very stable and can flow to other universality class when perturbations are introduced.

\subsection{Random qudit  graph state}
\label{sec:qudit}

In this section, we generalize the above idea to the random qudit graph state. We first briefly review the properties of qudit stabilizer state and then investigate the entanglement phase transition in the random qudit graph state by varying the measurement directions of the bulk qudits.

\subsubsection{Summary of qudit stabilizer state}\label{sec:quditStabilizer}
Qudit, similar to qubit, defines a $q$-dimensional Hilbert space spanned by a set of orthonormal basis $\mathcal{B}_q = \{\ket{j},j\in \mathbb{Z}_q\}$. Generalized qudit Pauli operators are defined as follows
\begin{equation}\label{eq:quditPauliDef}
Z_q = \sum_{\mu = 0}^{q- 1}\omega_q^\mu \ket{\mu}\bra{\mu}, \quad X_q = \sum_{\mu = 0}^{q -1} \ket{\mu+1}\bra{\mu},
\end{equation}
where $\omega_q = e^{2\pi i/q}$ is the $q$-root of unity. The extended commutation relation is
\begin{equation}
Z_q X_q = \omega_q X_q Z_q.
\end{equation}
From now on, we omit the $q$ index for brevity. 
The qudit Pauli group is $\mathcal{P} \cong \langle X,Z\rangle $, and the $N$-qudit Pauli group is generated by the $X$ and $Z$ operators of each qudit,
\begin{equation}\label{eq:NQuditPauliGroup}
    \mathcal{P}^{\otimes N } \cong\langle X_{1},X_{2},\ldots,X_{N},Z_{1},Z_{2},\ldots,Z_{N}\rangle.
\end{equation}
A group element $p_n \in \mathcal{P}_q^{\otimes N}$ can thus be written as 
\begin{equation}
    p_{n} = \omega^{c_{n}} \prod_{i = 1}^N X_{i}^{ a^{n}_{i}}Z_{i}^{b^{n}_{i}}
\end{equation}
where $a^{n}_{i}, b^{n}_{i},c_{n} \in \mathbb{Z}_q$.
When $q$ is a prime number larger than $2$, we can define a qudit stabilizer group $\mathcal{S} \subset \mathcal{P}_q^{\otimes N}$, which is generated by $N$ commuting
and independent generalized Pauli string operators $S_n$. They uniquely define a stabilizer state $|\psi\rangle$ satisfying $S_n|\psi\rangle=|\psi\rangle$. As a consequence, the full information of $|\psi\rangle$ can be conveniently stored in the stabilizer tableau which is a $N\times 2N$ matrix
\begin{equation}\label{eq:qdTableauRep.}
    T_q = [T_q^X,T_q^Z] 
\end{equation}
over finite field $q$. The $n$th row of $T_q^X$ and $T_q^Z$ describe $\{a^n_i \}$ and $\{b^n_i\}$ of $S_n$. For such a stabilizer state, the entanglement entropy of a subsystem A with $N_A$ qudits takes the form 
\begin{equation}\label{eq:quditEntropy}
    S_A = \operatorname{rank}_q (T_A) - N_A,
\end{equation}
where $T_A$ is the truncated stabilizer tableau of subsystem $A$, $N_A$ is the number of qudits in subsystem $A$, and $\operatorname{rank}_q$ denotes the rank over finite field $q$. The derivation of this result can be found in App.~\ref{sec:qudit_EE}\cite{fattal_entanglement_2004-1}. 

We define the controlled gate in qudit system, in a similar way to the controlled qubit gates. For example, the qudit Controlled-Phase (CP) gate operating on qudit $i$ and $j$ is defined as
\begin{equation}\label{eq:controlPhase}
CP_{ij} = \sum_{\mu=0}^{q - 1}\ket{\mu}_i\bra{\mu}_i\otimes Z_{j}^\mu = \sum_{\mu,\nu=0}^{q - 1}\omega^{\mu\nu}\ket{\mu}_i\bra{\mu}_i \otimes \ket{\nu}_j\bra{\nu}_j.
\end{equation}
When it operates conjugately on $X_{i}\slash Z_{i} \otimes I_j$, we have
\begin{equation}
    \begin{aligned}
    (CP_{ij}) Z_{i}\otimes I_j(CP_{ij})^\dagger &= Z_{i}\otimes I_j\\
    (CP_{ij}) X_{i}\otimes I_j(CP_{ij})^\dagger &= X_{i}\otimes Z_{j}.
    \end{aligned}
\end{equation}
Repeatedly applying $CP_{ij}$ gives
\begin{equation}\label{eq:CPMapping}
     (CP_{ij})^{k} X_{i}\otimes I_j \left[(CP_{ij}^\dagger )\right]^k = X_{i}\otimes Z_{j}^{k}.
\end{equation}

On $n$th qudit, there is a set of projection operators $P_n^{\lambda}$ with $\lambda\in\mathbb{Z}_q$
\begin{equation}
    P^\lambda_n =\frac{1}{q}\sum_{m=0}^{q-1} \omega^{\lambda m} O_n^m
\end{equation}
where $\lambda\in \mathbb{Z}_q$ for stabilizer $O_n=X_{n}^a Z_{n}^b$ with $a,b\in \mathbb{Z}_q$. We use the projector as the measurement, and the post-measurement state is another stabilizer state.

Qudit graph state, similar to qubit graph state, is also defined on a graph $G = \{V,E\}$, where qudits of local dimension $q$ live on vertices $V$. Starting from an initial state $|\psi\rangle = \ket{+_q}^{\otimes |V|}$ with $X_q \ket{+_q} = \ket{+_q}$, we apply CP gates to any pair of qudits $a,b\in V$ connected by edge $(i,j)\in E$. Different from $q=2$ graph state, $\langle CP_{ij} \rangle \cong \mathbb{Z}_q$, meaning that we can define the weight $w_{ij}\in\mathbb{Z}_q$ on edge $(a,b) \in E$ by applying $CP_{ij}$ $w_{ij}$ times. As shown in Eq.\eqref{eq:CPMapping}, $CP_{ij}^{w_{ij}}:X_{i}\otimes I_j \mapsto X_{i}\otimes Z_{j}^{w_{ij}}$. The $n$th generator of the qudit stabilizer group is thus
\begin{equation}
    P_{n} = X_{n}\prod_{m|(m,n)\in E} Z_{m}^{w_{mn}}, \qquad  w_{mn}\in\mathbb{Z}_q.
\end{equation}

The tableau representation of the stabilizer group $T_q = [T_q^X, T_q^Z]$ then takes the form $T_q^X = I$ and $T_q^Z = T_G^w$, where $T_G^w$ denotes the weighted adjacency matrix of graph $G = \{V,E\}$ with weights $w_{mn}\in\mathbb{Z}_q$ assigned to each $(m,n)\in E$,
\begin{equation}
T_G^w = \left\{
    \begin{matrix}
    w_{mn} & \quad (m,n)\in E\\
    0 & \text{otherwise}
    \end{matrix}\right. .
\end{equation}
Noting that $CP_{mn} = CP_{nm}$, we have $w_{mn} = w_{nm}$, meaning that $T_G^w = (T_G^w)^T$.

The entanglement entropy of subsystem $A \subset G$ is 
\begin{equation}\label{eq:quditEEAdjMat}
    S_A = \operatorname{rank}_q(T_{A \overline{A} }^w)
\end{equation}
where $\overline{A} = V\backslash A$, $\operatorname{rank}_q$ is the rank over finite field $q$, and $T_{A \overline{A} }^w$ is defined as a submatrix of $T_{G}^w$ shown below
\begin{align}
    T_G^w=\begin{pmatrix}
    T_{AA}^w & T_{A \overline{A} }^w\\
    T_{\overline{A} A}^w & T_{\overline{A} \overline{A}}^w \end{pmatrix}.
\end{align}
\subsubsection{Entanglement transition in the qudit random graph state}

We consider the same geometry as the qubit system, shown in Fig. \ref{fig:cir}, with randomly assigned weights $1\le w_{i,j}\le q - 1$ to each edge $(i,j)\in E \subset G$. On this random qudit graph state, random qudit-$X/Z$ measurements are performed in the bulk while qudits on the top surface are left un-measured. The probability of conducting $X$ measurement is denoted as $p_x$, whereas for $Z$ measurement the probability is $1 - p_x$. Similar to the qubit case, $X$ measurement tends to create new edges while $Z$ breaks the edges between the neighbors. The competition between them can lead to an entanglement phase transition.

\begin{figure}
\centering
\includegraphics[width=0.45\textwidth]{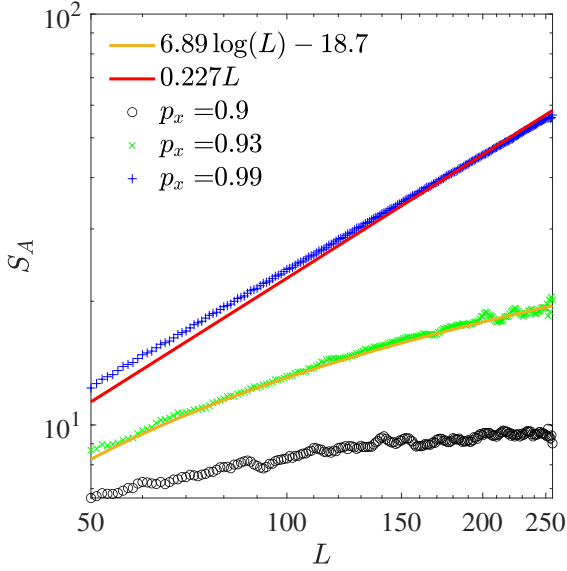}
\caption{Qudit $q = 3$ steady state entanglement entropy $S_A$ for various $p_x$. Here we take the square lattice system with $L_x = L_y = L$ and fix the ratio $L_A/L=1/4$. We plot $S_A$ as a function of $L$.}
\label{fig:q3EE}
\end{figure}

For a qudit system with $q = 3$, the entanglement phase transition is observed when tuning $p_x$. In Fig.\ref{fig:q3EE}, we fix the ratio of $L_A / L_x = 1/4$ and change the system size. When $p_x < p_x^c = 0.93$, $S_A$ saturates to a finite constant independent of system size, while at $p_x > p_x^c$, $S_A \sim L_x$. 

\begin{figure}
\centering
\includegraphics[width=0.45\textwidth]{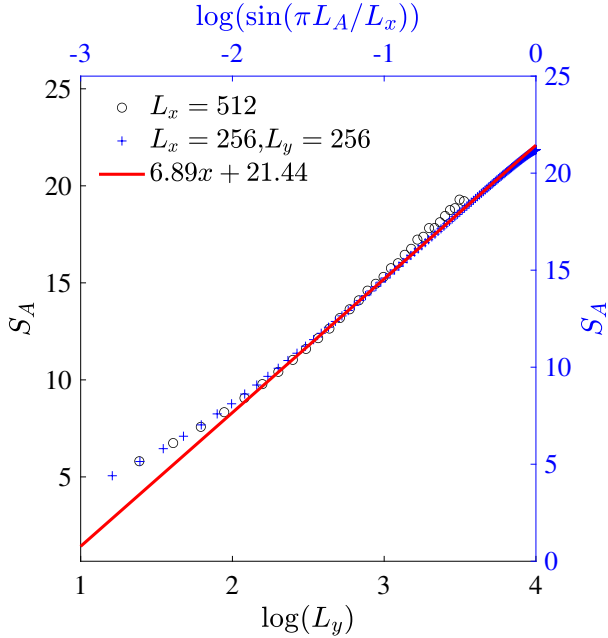}
\caption{Entanglement entropy growth of qudit $q = 3$ system at critical point $ p_x^c =  0.93$. Blue dots: $S_A$ vs. $\log(\sin(\pi L_A / L_x))$, where the system size is fixed as $L_x = L_y = 256$. Black dots: $S_A$ vs. $L_y$ with system size $L_A = L_x/2 = 256$, where $S_A$ scales linearly with $\log(L_y)$ when $L_y \ll L_x$. }
\label{fig:q3EEcrit}
\end{figure}

At critical point $p_x^c = 0.93$, $S_A$ scales logarithmically to the system size. More precisely, it satisfies Eq.\eqref{eq:EE} with $\alpha = 3.45$. This result is shown as blue dots in Fig. \ref{fig:q3EEcrit}.
Moreover, we calculate the entanglement entropy $S_A$ of half system with $L_A=L_x/2$ as a function of $L_y$. We find that $S_A$ grows linearly with $\log L_y$, the slope of which equals $2\alpha = 6.89$. The result is shown as the black dots in Fig. \ref{fig:q3EEcrit}.

\begin{figure}
\centering
\includegraphics[width=0.45\textwidth]{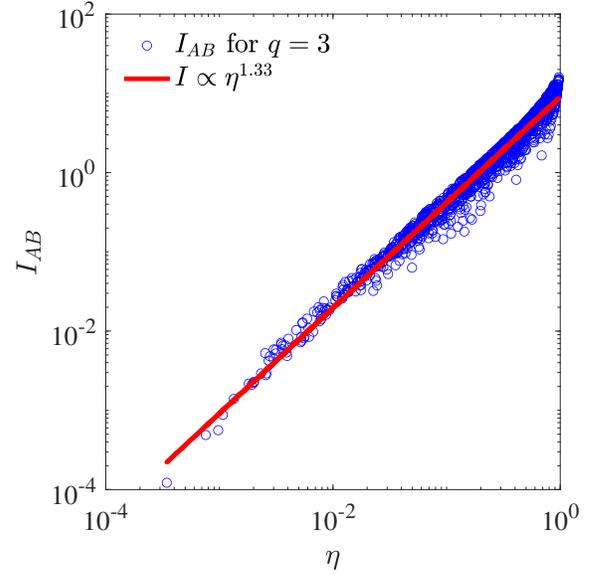}
\caption{ Mutual information $I_{AB}$ as a function of cross ratio $\eta$, at critical point $p_x^c = 0.93$ with $q = 3$.}
\label{fig:miq3}
\end{figure}

We also investigate the mutual information $I_{AB}$ as a function of cross ratio $\eta$ of two randomly chosen disjoint intervals $A = [x_1, x_2]$ and $B = [x_3, x_4]$ on the top surface as shown in Fig. \ref{fig:mtA}. At critical point $p_x = 0.93$, $I_{AB} \sim \eta^\Delta$ with $\Delta = 1.33$ as shown in Fig. \ref{fig:miq3}. 

\begin{figure}
\centering
\includegraphics[width=0.45\textwidth]{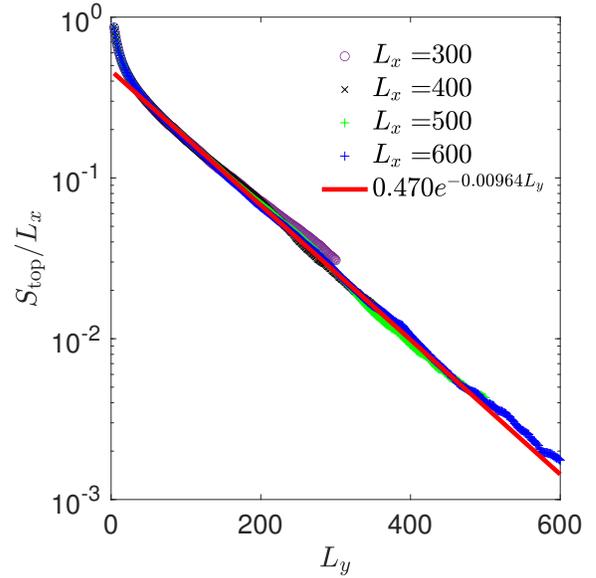}
\caption{The entanglement entropy $S_{\mbox{top}}$ between the top and bottom boundaries in the $q = 3$ qudit system, at $p_x^c = 0.93$. }
\label{fig:puriDq3Data}
\end{figure}

For the purification dynamics, we use the same set up shown in Fig. \ref{fig:puriDq2}. At the critical point, we find that the entanglement entropy decay exponentially as it is in the qubit system. The numerical result is shown in Fig.~\ref{fig:puriDq3Data}.

We also study the qudit system with other primer $q$ in the same way. We observe phase transitions in all of these systems and summarize the critical exponents in the Table \ref{tab:table3}. This result indicates that they belong to distinct universality classes for different $q$.

\begin{table*}
\caption{\label{tab:table3} critical exponents of random qudit graph state}
\begin{ruledtabular}
\begin{tabular}{cccccccc}
 Local Dim. $q$      & $2$    & $3$     & $5$   & $7$   & $23$ & $97$ & $997$ \\ \hline
 $p_c$               &$0.95$  & $0.93$  &$0.93$ &$0.925$&$0.92$&$0.92$& $0.92$\\
 $\alpha$            &$3.27$  & $3.45$  &$3.66$ &$3.48$ &$3.17$&$3.19$& $2.99$\\
 $\Delta$            &$1.05$  & $1.33$  &$1.34$ &$1.28$ &$1.36$&$1.36$& $1.36$
\end{tabular}
\end{ruledtabular}
\end{table*}

\section{Shallow circuit generated by random Clifford gates}
\label{sec: random_clifford}
\begin{figure}
\centering
\includegraphics[width=0.45\textwidth]{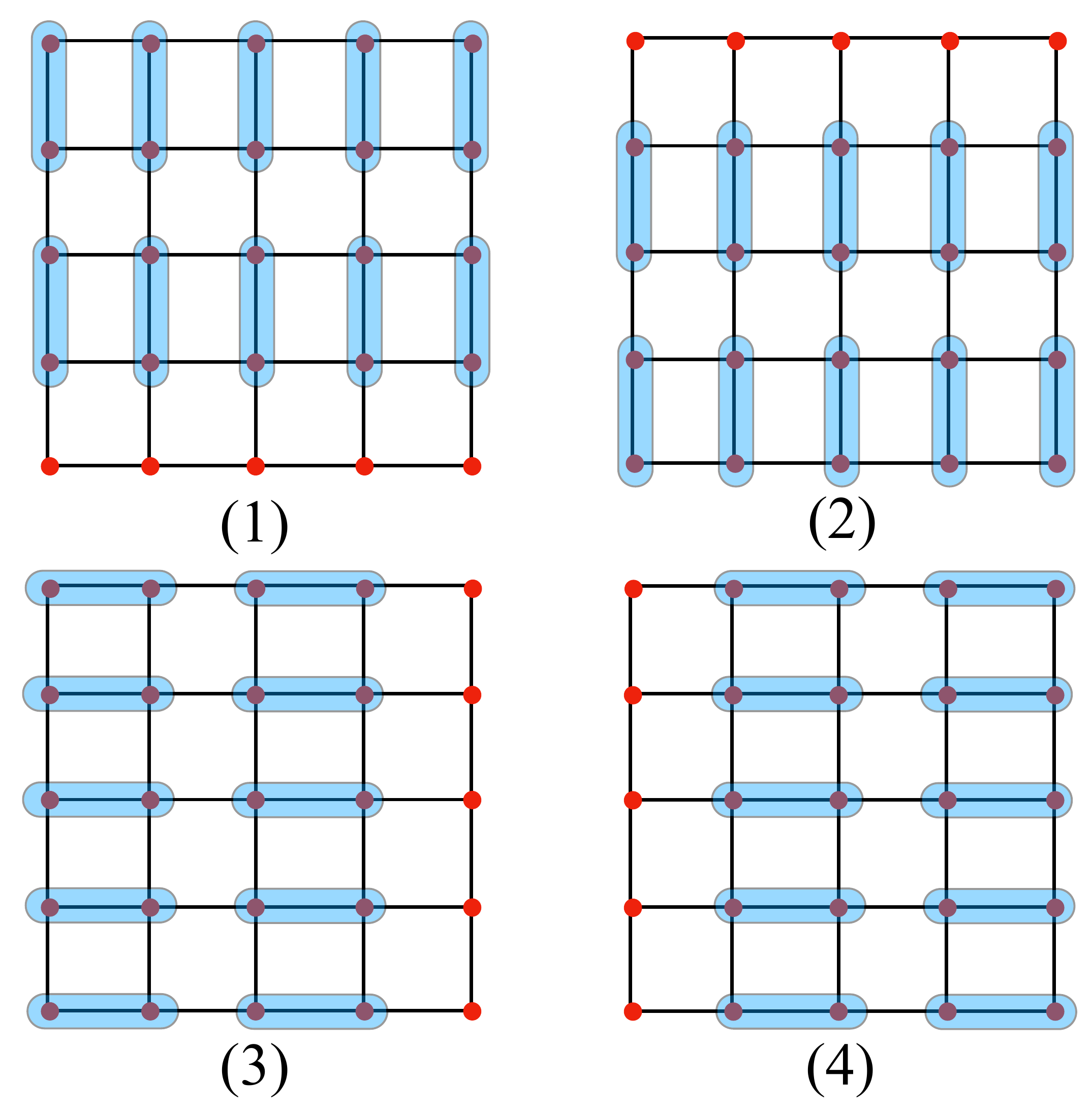}
\caption{The schematics for the unitary dynamics in one time step on the two dimensional square lattice. Each time step involves four successive layers of two-qubit gates with the pattern shown in (1)-(4). Here the qubits (red dots) are living on the vertices of the lattice and random two-qubit Clifford gates are denoted by the blue rounded rectangles. }
\label{fig:rd_schematics}
\end{figure}

So far we focus on the shallow circuit constructed of one layer of CZ/CP gates. In this section, we construct a shallow circuit composed of random two-qubit Clifford gates defined on the rectangular lattice. In each time step, we apply the gates along all the bonds in the square lattice. Different from CZ gates, the four gates acting on the same qubit may not commute with each other. Consequently, we split these gates into four layers shown in Fig.~\ref{fig:rd_schematics} and all the gates in one layer commute. 

By applying random shallow circuit with finite time steps $t$ on the product state $|\psi_0\rangle$ in Eq.\eqref{eq:polarized_x}, we obtain an area law entangled state $|\psi(t)\rangle$ defined on the rectangular lattice. We measure the bulk qubits and explore the potential entanglement transition for the one dimensional boundary state.

Here the random two-qubit Clifford gates are drawn uniformly from the two-qubit Clifford group. With this choice, the entanglement scaling for the boundary qubits is independent of the measurement direction of the bulk qubits. Therefore we simply take the projective measurement of the bulk qubits in the $Z$ direction. Numerically, we compute the entanglement entropy for the top boundary shown in Fig.~\ref{fig:cir} with sufficiently large $L_y$. When the time step $t=1$, the entanglement entropy has an area law scaling. On the other hand, when $t\geq 2$, the entanglement entropy has a volume law scaling. Similar measurement induced phase transitions have also been observed in the random tensor network \cite{yang2021entanglement,Vasseur2019} and in the random shallow circuit \cite{napp2020efficient,bao2022finite}. The latter has an interesting interpretation in terms of the output sampling complexity transition \cite{napp2020efficient}.

To design a continuous phase transition for the boundary qubits, we modify the above circuit slightly and introduce a tuning parameter $p\in[0,1]$. The two-qubit gate now becomes a random two-qubit Clifford gate with probability $p$ and is an identity operator with probability $1-p$. For this model, we expect that when $t$ takes finite value $\geq 2$, there exists an entanglement phase transition for the boundary qubit at finite $p_c$. Numerically, we take $t=2$ and observe that the critical point $p_c=0.744$. 

\begin{figure}
    \centering
    \subfigure[]{\label{fig:shallow_EE} \includegraphics[width=.42\textwidth]{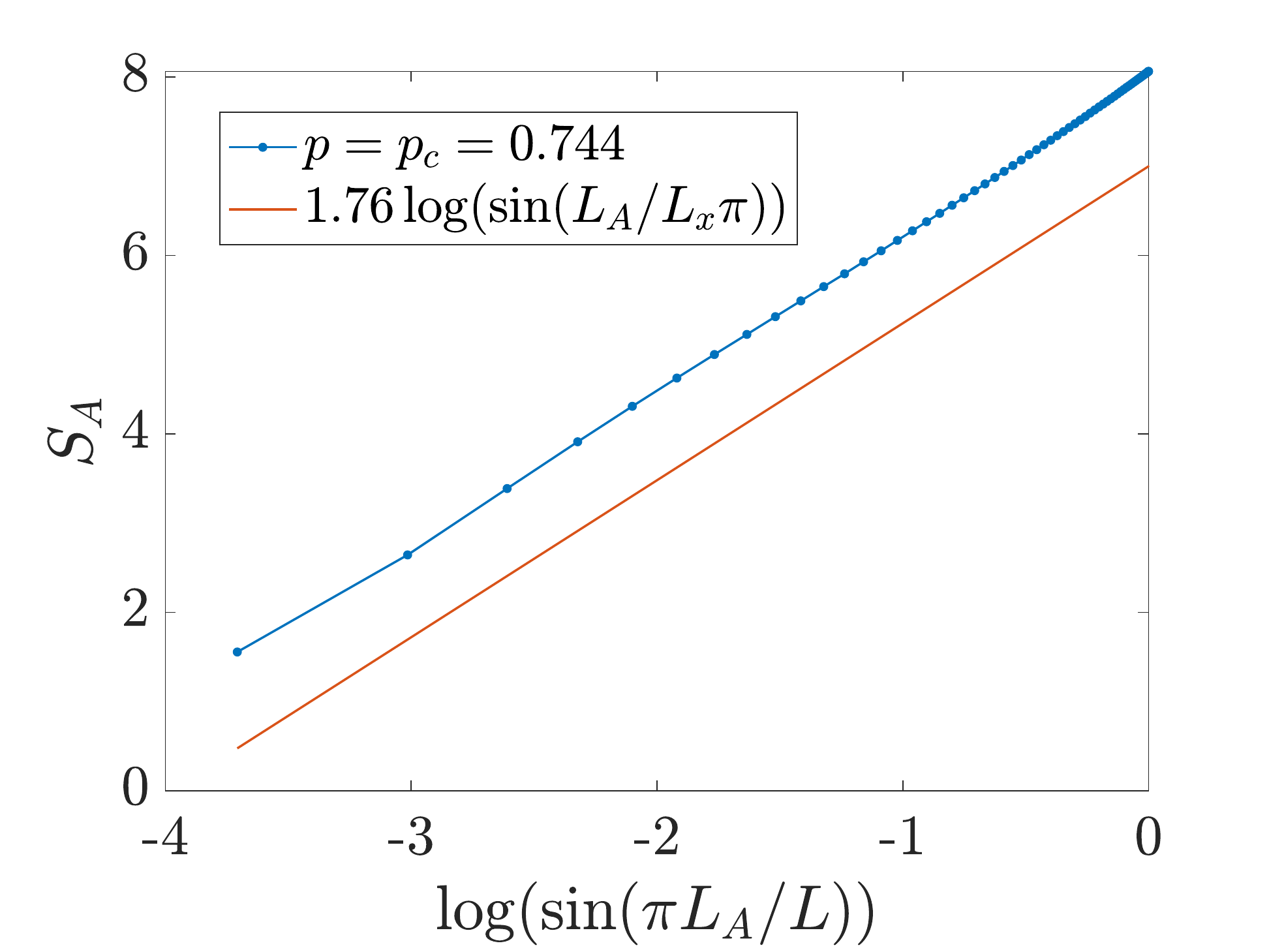}}
    \subfigure[]{\label{fig:shallow_MI} \includegraphics[width=.42\textwidth]{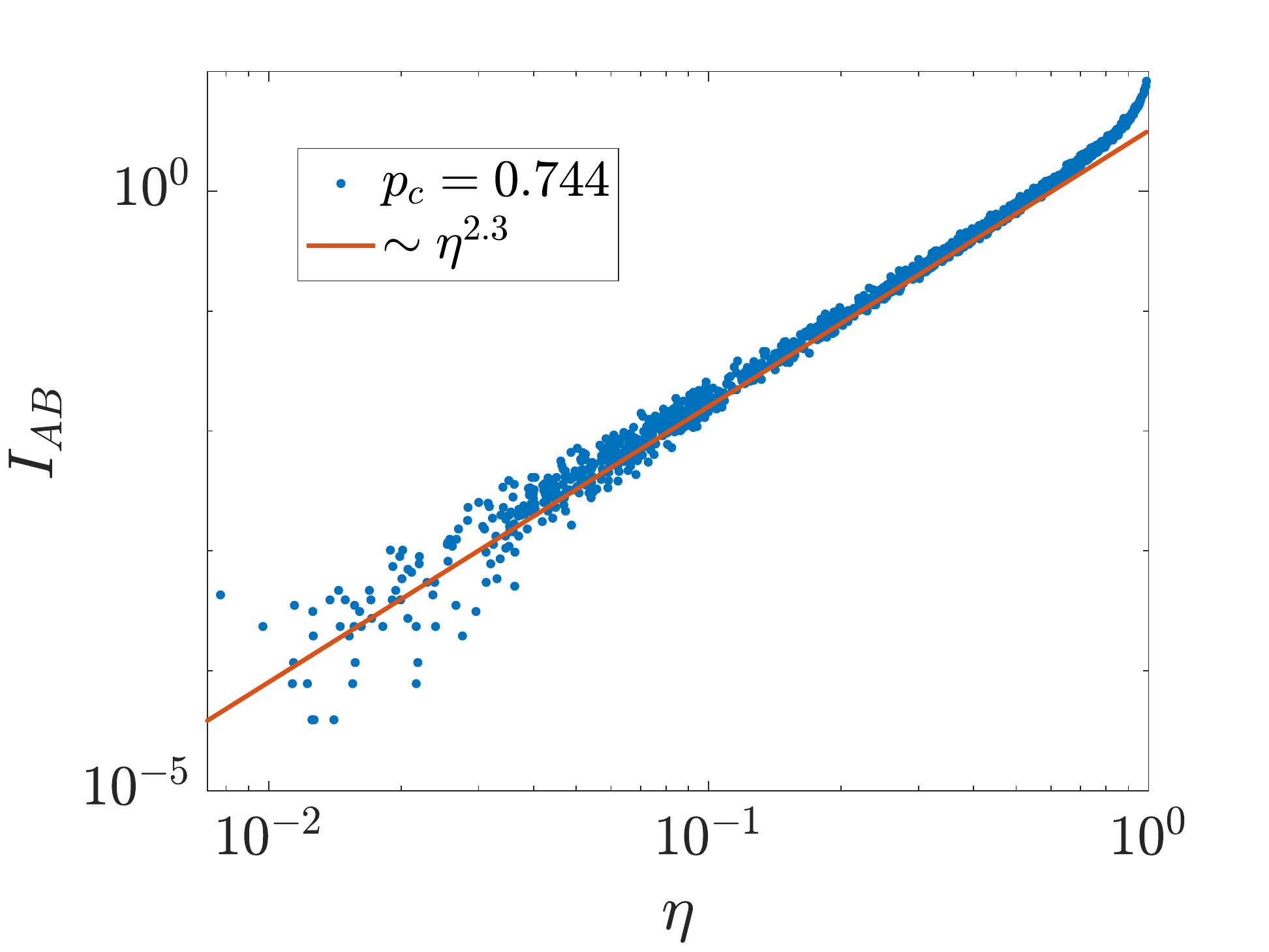}}
    \caption{The entanglement properties of the top boundary described in Fig.~\ref{fig:cir} at the critical point $p_c=0.744$. We take $L_x=256$ and $L_y=160$. The value of $L_y$ is large enough for $S_A$ to saturate. (a) $S_A$ is linearly proportional to  $\log (\sin(\pi L_A/L_x))$ and is consistent with the formula in Eq.\eqref{eq:EE}. (b) The mutual information $I_{AB}$ of two randomly chosen intervals as a function of  the cross ratio $\eta$ defined in Eq.\eqref{eq:eta}.}
    \label{fig:shallow}
\end{figure}

We investigate the entanglement scaling at the critical point. $S_A$ in the top boundary also takes the form in Eq.\eqref{eq:EE} with $\alpha=0.88$ (See Fig.~\ref{fig:shallow_EE}). We also compute the mutual information between two disjoint intervals and the results in Fig.~\ref{fig:shallow_MI} indicate that it is a function of cross ratio. In particular, $I_{AB}\sim \eta^{2.3}$ when $\eta\ll 1$. Again these results are obtained with a large $L_y$ so that the wave function of the top boundary reaches steady state.

\begin{figure}
    \centering
    \subfigure[]{\label{fig:shallow_puri_collapse} \includegraphics[width=.42\textwidth]{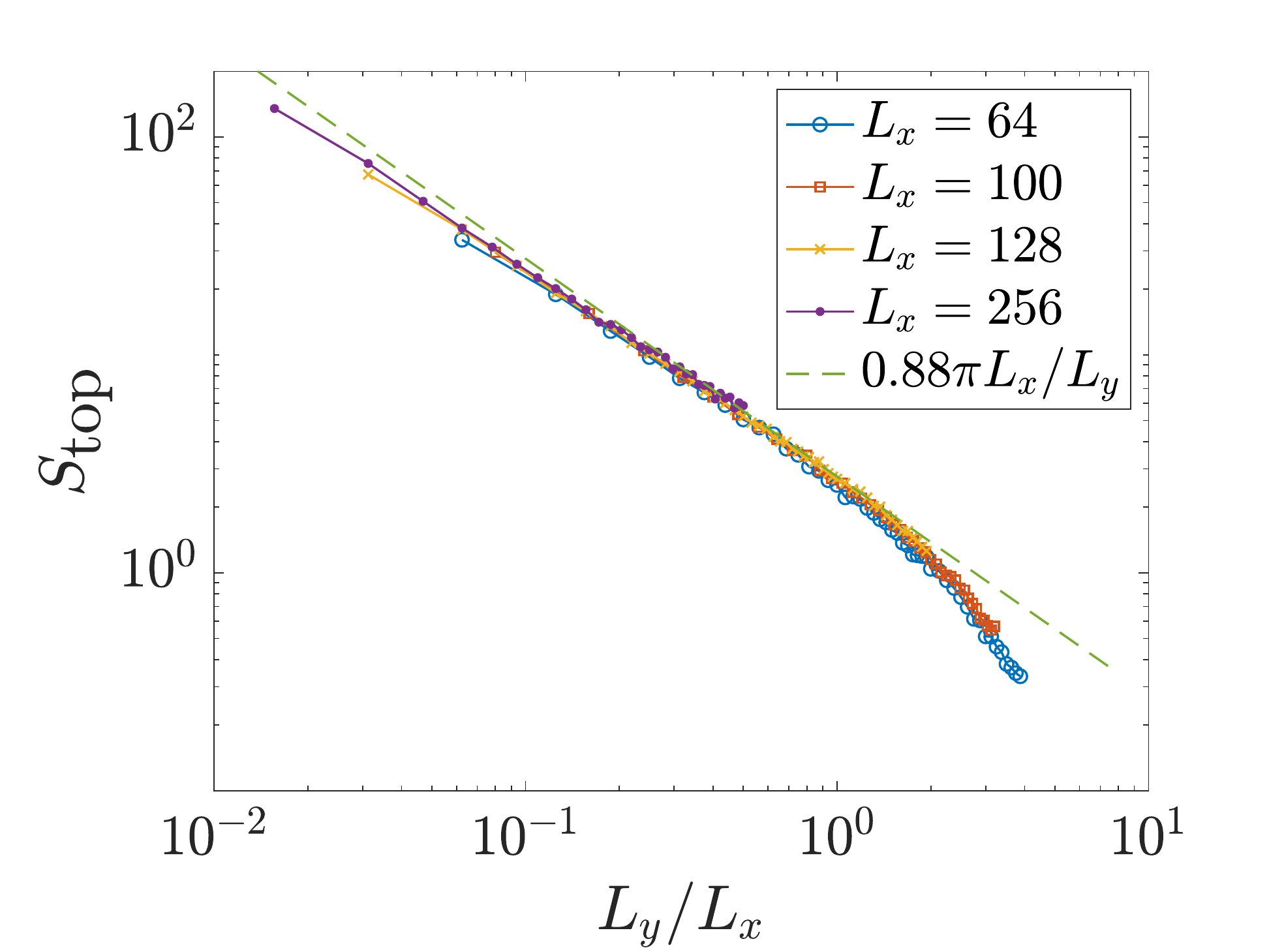}}
    \subfigure[]{\label{fig:shallow_puri_late} \includegraphics[width=.42\textwidth]{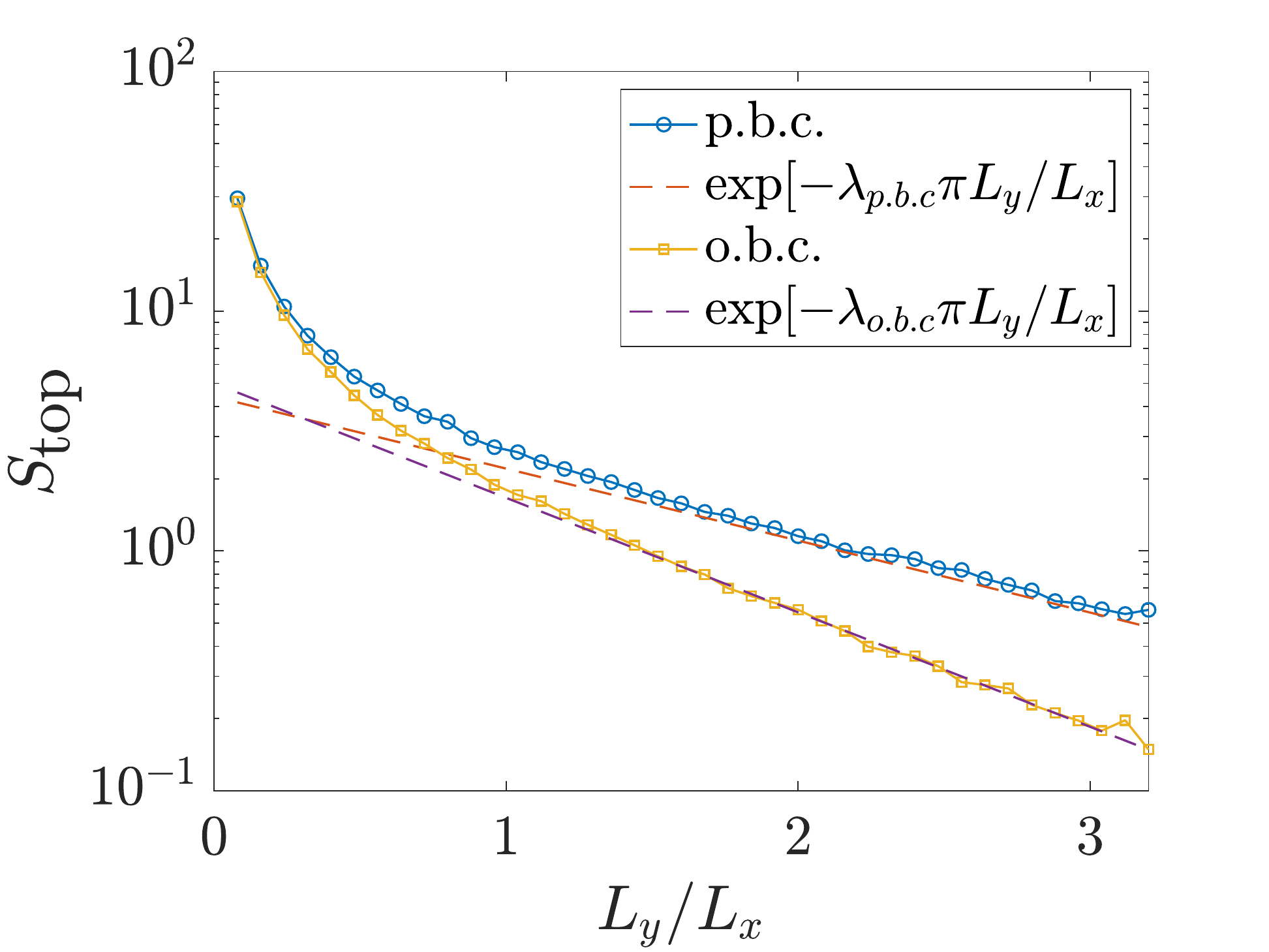}}
    \caption{The entanglement entropy $S_{\mbox{top}}$ between the top and bottom boundaries against the ratio $L_y/L_x$. (a) The data collapse of $S_{\mbox{top}}$ for different $L_x$ with periodic boundary condition (p.b.c.) along $x$ direction. (b) $S_{\mbox{top}}$ decays exponentially in $L_y/L_x$ when $L_y/L_x\gg 1$. The exponent $\lambda_{p.b.c}=0.22$ for p.b.c. and $\lambda_{o.b.c}=0.35$ for open boundary condition (o.b.c.). In the simulation, we take $L_x=100$.}
    \label{fig:shallow_puri}
\end{figure}

Next, we turn to the boundary condition described in Fig.~\ref{fig:puriDq2}. We vary both $L_x$ and $L_y$ and demonstrate that the entanglement entropy between the top and bottom boundaries $S_{\mbox{top}}= g(L_y/L_x)$ by performing data collapse in Fig.~\ref{fig:shallow_puri_collapse}. In particular, we observe that 
\begin{align}
    g(\tau)= \left\{\begin{array}{cl}
\frac{\alpha \pi }{\tau}, &\tau\ll 1\\
a\exp(-\lambda\pi\tau),& \tau\gg 1\end{array}\right.,
\end{align}
where $\tau=L_y/L_x$ and $a$ is a non-universal number. Both the power law decay with small $\tau$ (Fig.~\ref{fig:shallow_puri_collapse}) and the exponential decay with large $\tau$ (Fig.~\ref{fig:shallow_puri_late}) are also observed in random tensor networks and the purification dynamics in the hybrid circuits\cite{li2020conformal,gullans2020dynamical,yang2021entanglement}. Such scaling can be understood by assuming that the entanglement entropy can be mapped to the free energy of a statistical mechanical model discussed in Sec.~\ref{sec:haar}. The $\alpha$ and $\lambda$ are universal exponents of the critical statistical mechanical model with two dimensional conformal symmetry. 

The above analysis can be generalized to the Clifford qudit circuits with $q>2$. Since the transitions will be very similar, we will not study them numerically in this paper. One small difference we observe is that when $q>2$, if the time step $t=1$, the boundary state is volume law entangled at $p=1$. This indicates that for large $q$, we only need to take a shallow circuit with four layers of gates described in Fig.~\ref{fig:rd_schematics}. By decreasing $p$, there exists a continuous entanglement phase transition in it. In the next section, we will consider a similar random Haar  circuit with four layers of gates and provide an interpretation of this entanglement transition.

\section{Transition in random Haar  circuit}
\label{sec:haar}

%% consider Haar gate, specify geometry
In this section, we consider the same entanglement transition in a random Haar circuit. The circuit geometry and sequences of gates to apply remain the same as in Fig.~\ref{fig:rd_schematics}, but the gate is taken as independent random Haar  unitary of dimension $q^2 \times q^2$. The choice of gate ensures that the dynamics is strongly chaotic. When the depth of the circuit reaches order $L$, the resulting 2d final state is a random state with almost maximal entanglement for any partition\cite{page_average_1993}. In this case, measuring the bulk qudits creates a random state on the boundary, which has volume law entanglement. On the other hand, in a shallow circuit, the entanglement entropy for a two dimensional subsystem obeys area law. By measuring the bulk degrees of freedom, we explore the possible entanglement phase transition of the 1d boundary state.

Thanks to the random matrix theory, the random averaging quantities of random circuit can be mapped to a statistical mechanical models of emergent spins\cite{hayden2016holographic,Nahum_2017,zhou2019emergent,khemani_operator_2018,chan_solution_2018,von_keyserlingk_operator_2018,bao2020theory}. The time dependent R\'enyi entropy is the free energy of the spin model. The quantity we consider is the quasi-entropy 
\begin{equation}
\tilde{S}_2 = - \ln \frac{\overline{Z_A}}{\overline{Z_{\emptyset}}}
\end{equation}
where
\begin{equation}
\overline{Z_A} = \tr( \tilde{\rho}_A^2 ) \quad Z_{\emptyset} = \tr( \tilde{\rho}^2 ) 
\end{equation}
with $\tilde{\rho}$ the unnormalized density matrix and $\tilde{\rho}_A$ the unnormalized reduced density matrix. There are different conventions/formalisms to carry out the random averaging\cite{Nahum_2017,zhou2019emergent,bao2020theory,jian2019measurementinduced,napp2020efficient,fan2020self}. We adopt the convention of Ref.~\onlinecite{napp2020efficient} for its similarity with our circuit geometry. The mapping is briefly reviewed in Sec.~\ref{subsubsec:napp} via a similar 2d system example in Ref.~\onlinecite{napp2020efficient} and then is further carried out for our model in Sec.~\ref{subsubsec:ourmodel}. More details are spared in App.~\ref{app:haar}. In summary, the effective spins live on the bonds (part of the bonds, after intergarting out some of the spins) and they interact with their spatial neighbors with ferromagnetic interactions. In the case where unitary gates are applied with certain probability, the vacant gates will appear as broken bonds in the Ising model. In the large $q$ limit, the bulk transition can be described by a random bond Ising model, which also entails the associated boundary transition in question. We expect this transition to persist to finite $q$ and to architectures different from Fig.~\ref{fig:rd_schematics}, though the nature of the generic transition requires further investigation. 

\subsection{The spin model with only two-body interaction}
\label{subsubsec:napp}

\begin{figure}[h]
\centering
\subfigure[]{
  \label{fig:napp_gate}	
  \includegraphics[width=0.8\columnwidth]{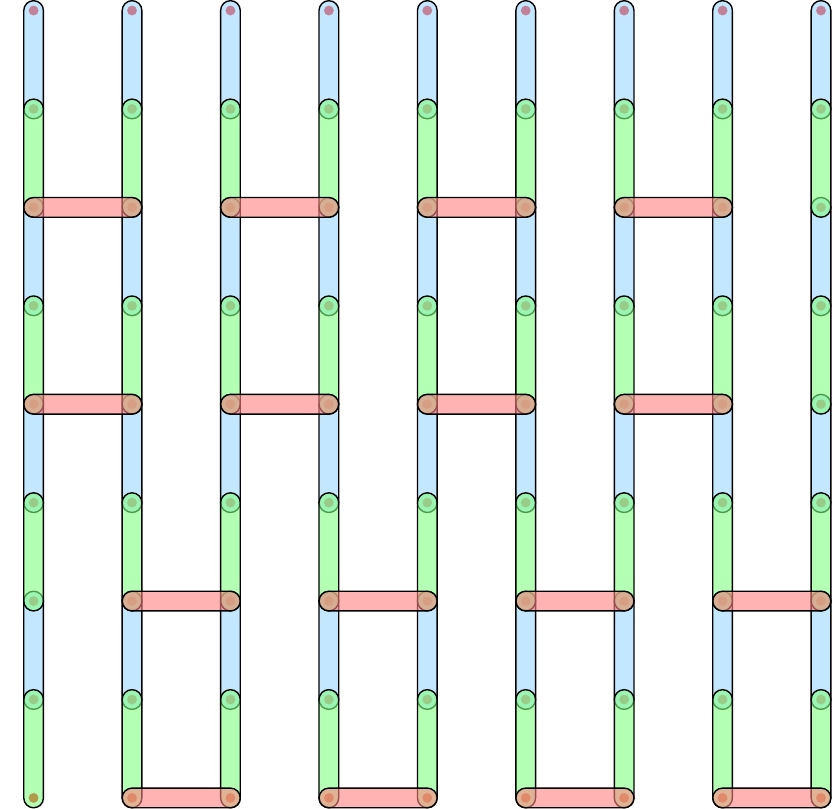}
}
\subfigure[]{
  \label{fig:napp}	
  \includegraphics[width=0.8\columnwidth]{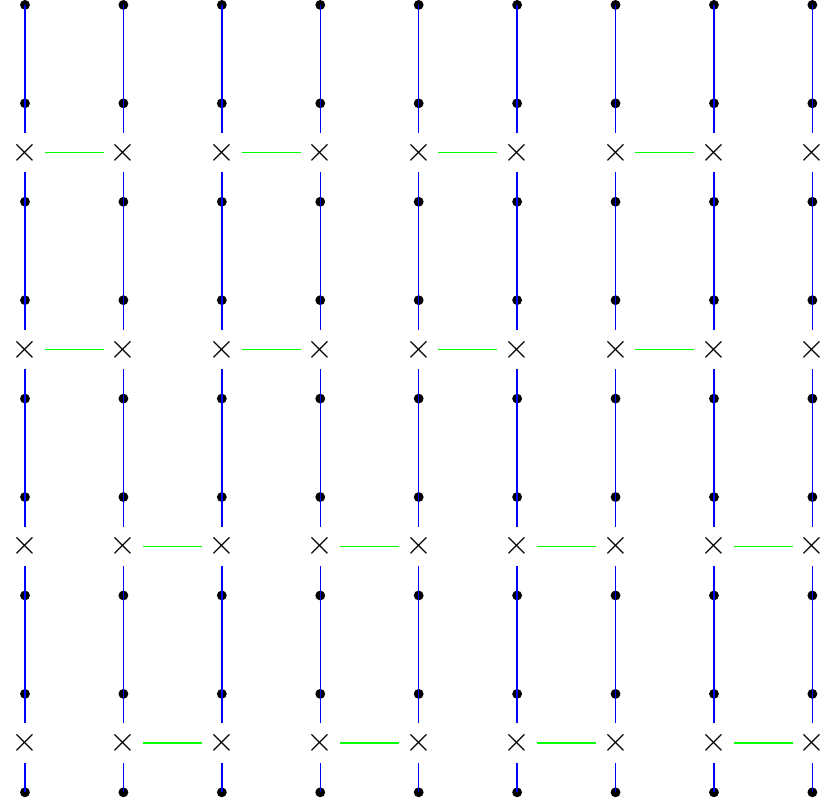}
}
\caption{The 2d model in Ref.~\onlinecite{napp2020efficient}. (a) The first (blue), second (green) and third layer (red) of gates applied on the lattice (black dots). (b) The effective spin models. The spin (cross) lives on the even vertical bonds. They interact vertically by integrating out the spins on the odd bonds. They also interact horizontally via integrating out the spins on the third layer of gates. }
\label{fig:napp_spin}
\end{figure}

In this subsection, we review the mapping to a spin model via a 2d circuit example in Ref.~\onlinecite{napp2020efficient}, whose setup is reproduced in Fig.~\ref{fig:napp_gate}. Since our setup is very similar to Ref.~\onlinecite{napp2020efficient}, we will not repeat the calculation in the main text, but only to review the descriptions of the spin variables and their interactions only. Some details are deferred to App.~\ref{app:haar}. 

As mentioned earlier, we aim to compute the quasi-entropy, which is written as the ratio of two partition functions in the random average. In both the partition function, $Z_A$ and $Z$, the time evolved density matrix appears twice. Thus each random unitary gate $u$ appears twice and so does its complex conjugation. A random average performed over $u\otimes u^* \otimes u \otimes u^*$ produces a linear combinations of permutation elements\cite{collins_integration_2006,collins_moments_2002,brouwer_diagrammatic_1996}(also see App.~\ref{app:haar}). Physically they represent different ways to pair the unitary $u$ and its complex conjugate $u^*$\cite{zhou_entanglement_2020}. In our particular setup with only second moment of $\tilde{\rho}$, the two permutations are $\I$ and $(12)$:
\begin{align}
   \I &: \quad \contraction[1.5ex]{}{u}{\otimes }{u^*} 
   \contraction[1.5ex]{u \otimes u^* \otimes }{u}{\otimes }{u^*} 
   u\otimes u^* \otimes u \otimes u^* \quad (s = 1)\\
   (12) &: \quad \contraction[1.5ex]{}{u}{\otimes u^* \otimes u \otimes }{u^*} 
   \contraction[1.5ex]{u \otimes }{u^*}{\otimes}{u} 
   u\otimes u^* \otimes u \otimes u^* \quad (s = -1)
\end{align}
These are the two Ising spins in our problem. We may just call them $s = \pm 1$. 

Evaluating either $\overline{Z_A}$ or $\overline{Z}$ will result in a tensor network/graph of these permutation spins, whose legs without internal contractions represent the finial state. Apart from the qudits to be measured, the legs on the boundary are contracted with boundary permutation spins. In $\overline{Z}$ they are contracted uniformly with the $\I$ spins, while in $\overline{Z_A}$, the legs in boundary region A contract with $(12)$ and its complement with $\I$. The later is a domain wall boundary condition, which probes the bulk property by exciting a domain wall upon ``vacuum". The difference between the free energies $-\ln \overline{Z_A}$ and $-\ln \overline{Z}$ is the quasi-entropy.  Since the boundary entanglement phase transition is a manifestation of the bulk property, we can directly look at the bulk spin model and see if there is a bulk phase transition.

  For the circuit described in Fig.~\ref{fig:napp_gate}, after the application of the first two layers of horizontal gates and integrating out some of the spin variables,  the remaining spins are placed at the even vertical bonds. In Fig.~\ref{fig:napp}, we label the lattice sites by black dots and spins as crosses. The blue {horizontal} line between the crosses is the ferromagnetic interaction. 

  In Fig.~\ref{fig:napp_gate}, a third layer of gate is applied on a fraction of horizontal bonds\cite{napp2020efficient}. The place to act on these gates are specifically designed such that  in the corresponding spin model,  one spin participates only in one horizontal ferromagnetic Ising interaction (see Fig.~\ref{fig:napp}). Thus one can write down a ferromagnetic Ising model
\begin{equation}
H = - \sum_{\langle i j\rangle } J_{\rm vert} s_{i} s_j - \sum_{\langle i j\rangle } J_{\rm horiz} s_{i} s_j 
\end{equation}
where $J_{\rm vert}= \frac{1}{2} \ln \frac{q^2 + 1}{2q}$ and $J_{\rm horiz} = \frac{1}{2} \ln \frac{1 + 2q + 4q^2 + 2q^3 + q^4}{2q(1 + q + q^2)}$. The interactions are only turned on the blue and green lines in Fig.~\ref{fig:napp}. Ref.~\onlinecite{napp2020efficient} then argues about the bulk transition by tuning the value of $q$ from small to large. When $q = 2$, both $J_{\rm vert}$ and $J_{\rm horiz}$ are below the critical point of 2d Ising model, indicating that the system is in the disordered phase and the quasi-entropy obeys an area law. When $q$ is large, the system is in the order phase, and the quasi-entropy obeys a volume law.

In this model, the transition is not continuous since physically $q$ can only take integer values. To study the continuous phase transition considered in this paper, we fix $q$ to be a large value, so that it begins with a volume law entangled phase which corresponds to the ordered phase in the Ising model. Then we modify the rule such that we apply the gate in the first and third layers with probability $p$. In this case when there is a vacant gate in the first (third) layer, the corresponding vertical (horizontal) interaction on that bond will be zero. This is a random bond Ising model which undergoes a continuous phase transition by tuning $p$. When $p$ is large, this model is in the disordered phase. Therefore the boundary state entanglement transition can then be identified as the order-disorder transition by varying $p$.

\subsection{The transition in the four-layer geometry}
\label{subsubsec:ourmodel}

\begin{figure}[h]
\centering
  \includegraphics[width=0.8\columnwidth]{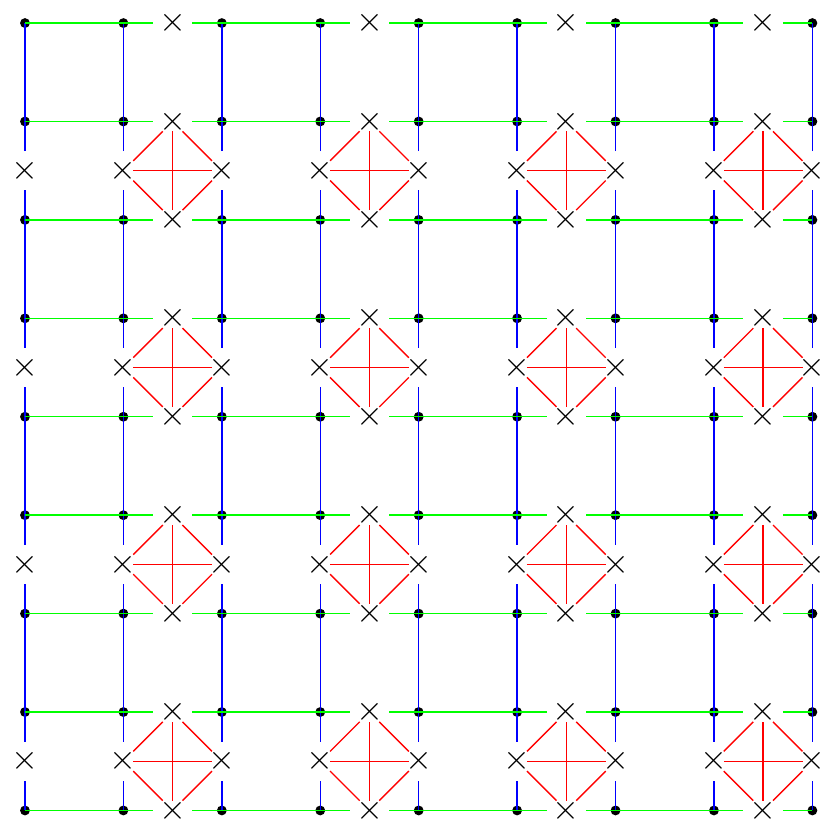}
\caption{Spin interactions in our model in Fig.~\ref{fig:rd_schematics}. The first (last) two layers of gates generate spins on even vertical (horizontal) bonds, and they interact vertically (horizontally) with their nearest neighbors. There are two-body and four-body interactions within a plaquette where the vertical and horizontal spins meet. A vacant gate in the first (and forth layer) means the absence of a blue (green) interaction. }
\label{fig:our_work}
\end{figure}

%% our work, construct a 2 and 4 spin ferromagnetic interacting model, when p = 0
The geometry we follow in this work has four layers of unitaries. We view the quantity as the overlap of the first two and last two layers (See the circuit in Fig.~\ref{fig:rd_schematics}). Similar to the previous subsection, the random average of the first two layers generates spins on the even vertical bonds. Their nearest neighbor vertical interaction strength is $J_{\rm vert}$. Going through the same reduction procedures of the first two layers,  the last two layers can further generate spins on the even horizontal bonds. These spins interact with their horizontal nearest neighbor also with strength $J_{\rm vert}$ (note not $J_{\rm horiz}$). After evaluating the overlap, four spins that meet in a plaquette have 4-body and 2-body interactions (in large $q$ limit), see Fig.~\ref{fig:our_work}. The effective Hamiltonian reads
\begin{equation}
\begin{aligned}
  H &= - \sum_{\langle ij \rangle \in | \text{or} - } J s_i s_j  - \sum_{\langle  i,j \rangle \in \square } J_{12} s_i s_j \\
  &- \sum_{\langle \langle   i,j \rangle \rangle \in \square } J_{13} s_i s_j  - \sum_{i,j,k,l \in \square }J_{1234}  s_i s_j s_k s_l .
\end{aligned}
\end{equation}
The first term has $J = J_{\rm vert}$. It represents the vertical and horizontal nearest neighbor interactions. Note that in Fig.~\ref{fig:our_work}, spins on the first row only interact horizontally, and spins on the first column only interact vertically. These two sublattices of spins interact together through the interaction in a plaquette. Here $J_{12}$, $J_{13}$ are 2-body ferromagnetic interaction strength for nearest neighbor and next nearest neighbor in a plaquette. $J_{1234}$ is the 4-body interaction strength. They are all order $\ln q$ in the lowest order expansion. The derivation of this Hamiltonian can be found in App.~\ref{app:haar}.

We assume the probability to apply a gate is $p$. For simplicity we only consider possible vacant gates in the first and fourth layers. In this case, we can show that a vacant gate amounts to setting the vertical or horizontal interaction $J$ to be $0$ at that bond. Thus we obtain a random bond Ising model with 4-body interactions in the plaquette. 

There are several simple limits. When we take the $q = \infty$ limit, all the interaction strength is infinite. The phase transition reduces to a geometric transition and will occur at the bond percolation critical point\cite{skinner2019measurement}.

When $q$ is finite but large, the interactions in the plaquette is stronger than the vertical and horizontal Ising interactions. Taking an approximation that the plaquette interactions are infinitely stronger, we can assume all the spins inside a plaquette to align {in the same direction}. We can then treat these four spins in a plaquette as a single Ising spin and end up with a random bond Ising model
\begin{equation}
H_{\text{rand bond}} = - \sum_{\langle ij \rangle  } J_{ij} s_i s_j 
\end{equation}
where $J_{ij} = 2 J_{\rm vert}$ with probability $p$ and $0$ otherwise. This is the standard random bond Ising model on a square lattice with nearest neighbor interactions. When the plaquette interaction is not much stronger, the model is more complicated. But we speculate that this model still undergoes a similar order-disorder phase transition as we vary $p$.

\section{Discussion and Conclusion}
\label{sec:conclusion}
In this paper, we explore measurement induced entanglement phase transition in two dimensional shallow circuits. Specifically, we prepare two dimensional resource states generated by shallow circuits composed of either one layer of controlled phase gates or the random Clifford/Haar gates. By performing local measurement for the bulk qubits/qubits, we show that there could exist continuous entanglement phase transition between an area law phase and a volume law phase for the one dimensional boundary qubits/qudits.

This measurement induced transition may not be limited to the shallow circuits. We could consider other area law entangled state, such as the critical states and topological phases which cannot be obtained by applying a shallow circuit on a trivial product state. It would be interesting to explore if there is an interesting entanglement structure for the boundary state by monitoring the bulk degrees of freedom.

Observing the entanglement phase transition in the noisy near-term quantum computer is an outstanding problem. For the hybrid circuit protocol, it requires a deep circuit with repeated measurement, which is difficult to realize on the current noisy devices. Our protocol has similar physics and yet needs high overhead. Since it requires only a shallow circuit and one layer of measurements at the end, it has the advantage over the hybrid circuit in terms of experimental realization. Recently, there is a proposal of preparing topological phase in Rydberg atoms by making single qubit measurements for a fraction of qubits\cite{tantivasadakarn2022longrange,verresen2022efficiently}. It would be interesting to use similar method to realize our protocol in the near-term devices.  

\acknowledgements
 We acknowledge the helpful discussions with
Chao-Yang Lu's group on the realization of shallow circuits and entanglement measurement on the  superconducting qubit platform. 

\appendix
\section{Entanglement entropy of qudit stabilizer state}
\label{sec:qudit_EE}
In this section we provide a detailed derivation of the qudit entanglement entropy formula Eq.\eqref{eq:quditEntropy} in the language of tableau representation of stabilizer group $\mathcal{S}$
\begin{equation*}
    S_A = \operatorname{rank}_q T_q^A - N_A
\end{equation*}
and the entanglement entropy formula Eq.\eqref{eq:quditEEAdjMat} 
\begin{equation}
    S_A = \operatorname{rank}_q T^w_{A \overline{A}}. 
\end{equation}

The qudit stabilizer group $\mathcal{S}$ of a quantum state $\ket{\psi}$ in Hilbert space $\mathcal{H}$ by definition gives
\begin{equation}
    s\ket{\psi} = \ket{\psi}, \quad \forall s \in \mathcal{S}
\end{equation}
where $\mathcal{S}$ is a subgroup of the $N$-qudit Pauli group defined in Eq.\eqref{eq:NQuditPauliGroup} and the Hilbert space $\mathcal{H}$ is spanned by a set of orthonormal basis $\mathcal{B}_q = \{\ket{\mu},\mu\in \mathbb{Z}_q\}$.

In other words, $\ket{\psi}$ defines a uni-dimensional invariant subspace $\mathcal{H}_\mathcal{S} = \{\ket{\psi}\} \subset \mathcal{H}$ of the stabilizer group $\mathcal{S}$. We may thus write down the rank-$1$ projection operator $\Pi_{\mathcal{S}} : \mathcal{H} \to \mathcal{H}_{\mathcal{S}}$
\begin{equation}
    \Pi_{\mathcal{S}} = \ket{\psi}\bra{\psi}.
\end{equation}

On the other hand, we can construct the projection operator $\Pi_\mathcal{S}$ directly from the representation of group $\mathcal{S}$. Since group $\mathcal S$ acts trivially on $\ket{\psi}$, it defines the trivial representation of $\mathcal{S}$, meaning that its character $\chi_{D_{\ket{\psi}}(s)} = 1 $ and dimension of the representation $n_{\ket\psi} = 1$ for all $s \in \mathcal{S}$. The projection operator $\Pi_\mathcal{S}$ is therefore written as
\begin{equation}
    \Pi_\mathcal{S} =\frac{n_{\ket{\psi}}}{|\mathcal{S}|} \sum_{s\in \mathcal{S}} \chi^*_{\ket{\psi}}D(s) = \frac{1}{|\mathcal{S}|}\sum_{s \in \mathcal{S}} D(s)
\end{equation}
where $D(s)$ denotes the representation defined in $\mathcal{H}$ and it takes the form
\begin{equation}
    D(s) = \omega^{c_s} \prod_{i = 1}^N X_{i}^{a_{i}}Z_{i}^{b_{i}}
\end{equation}
where $a^{n}_{i}, b^{n}_{i},c^{n} \in \mathbb{Z}_q$, $\omega_q$ denotes the $q$-root of unity, and $X_{i}/Z_{i}$'s are the generalized Pauli operators defined in Eq.\eqref{eq:quditPauliDef}. One more thing worth noticing here is that when $a^n_i = b^n_i = 0$ we have $D(s) = I_{q^N}$, the $q^N$ dimensional identity matrix. Since $\Pi_{\mathcal{S}}$ is a rank-$1$ projection operator, taking trace on both sides we have 
\begin{equation}
    \operatorname{Tr}\Pi_{\mathcal{S}} = \frac{1}{|\mathcal{S}|}\sum_{s \in \mathcal{S}} \operatorname{Tr} D(s) = \frac{q^N}{|\mathcal{S}|} = 1
\end{equation}
meaning that $|\mathcal{S}| = q^N$.

We now have 
\begin{equation}
    \rho = \ket{\psi}\bra{\psi} = \frac{1}{q^N}\sum_{s \in \mathcal{S}} D(s)
\end{equation}
where $\rho$ is the density operator of the whole system. Let $\rho_A$ be the reduced density operator of subsystem $A$, defined as $\rho_A = \operatorname{Tr}_{\overline{A}} \rho$, where $\operatorname{Tr}_{\overline{A}}$ denotes the partial trace over the complement of the total system $V$, $\overline{A} = V\backslash A$, and thus
\begin{equation}\label{eq:reducedDensity}
    \rho_A = \operatorname{Tr}_{\overline{A}}\rho = \frac{1}{q^N}\sum_{s \in \mathcal{S}} \operatorname{Tr}_{\overline{A}} D(s).
\end{equation}
Noting that $\operatorname{Tr} (X_q^m Z_q^n) = q \delta_{m,0}\delta_{n,0}$ for $a,b \in \mathbb{Z}_q$,
\begin{equation}
   \sum_{s \in \mathcal{S}} \operatorname{Tr}_{\overline{A}} D(s) = q^{|\overline{A}|}\sum_{s_A \in \mathcal{S_A}}D'(s_A)
\end{equation}
where $\mathcal{S}_A \subset \mathcal{S}$, and $D'(s_A)$ is a submatrix of $D(s)$ taking only the entries of system $A$, that is 
\begin{equation}
     D'(s_A) = \omega_q^{c_{sA}}\prod_{i \in A} X_{i}^{ a^{n}_{i}}Z_{i}^{b^{n}_{i}}.
\end{equation}

The reduced density operator $\rho_A$ in Eq.\eqref{eq:reducedDensity} takes the form
\begin{equation}
     \rho_A = \operatorname{Tr}_{\overline{A}}\rho = q^{-|A|}\sum_{s_A \in \mathcal{S_A}}D'(s_A).
\end{equation}
One quantity that interests us is the trace of $\rho_A^\alpha$
\begin{equation}
\begin{aligned}
     \operatorname{Tr}\rho_A^\alpha &= \operatorname{Tr} q^{-\alpha |A|} \prod_{i = 1}^\alpha \sum_{s_{A,i}\in \mathcal{S}_A} D'(s_{A,\alpha}) \\
     &=  q^{-\alpha |A|} |\mathcal{S}_{A}|^{\alpha-1}\sum_{s_A \in \mathcal{S}_A}\operatorname{Tr}D'(s_{A,\alpha}) \\
     &= \left(\frac{|\mathcal{S}_A|}{q^{|A|}}\right)^{\alpha - 1}.
\end{aligned}
\end{equation}
We now calculate $\alpha$-R\'enyi entropy of $A \subset V$
\begin{equation}
    S_A^\alpha = \frac{1}{1 - \alpha} \log_q \operatorname{Tr} \rho_{A}^\alpha = |A| - \log_q|\mathcal{S}_A|.
\end{equation}
Let $\mathcal{G}(\mathcal{S}_A) \subset \mathcal{P}_q^{\otimes N}$ denote the set of generators of the quotient group $\mathcal{S}_A = \mathcal{S}/\mathcal{S}_{\overline{A}}$. The number of its element is $\log_q|\mathcal{S}_A|$. So the $\alpha$-R\'enyi Entropy takes the same form as the qubit system
\begin{equation}
    S_A^\alpha = |A| - |\mathcal{G}(\mathcal S_A)|.
\end{equation}
Now, we follow the same step as in Ref.~\onlinecite{Li_2019}, and consider a projection operator $\operatorname{proj}_A$
\begin{equation*}
    \begin{aligned}
    \operatorname{proj}_A : \mathbb{Z}_q^{N,2N} &\to \mathbb{Z}_q^{N,2|A|}\\
    T_q &\mapsto T_q^A
    \end{aligned}.
\end{equation*}
where $T_q$ is the tableau representation of the qudit stabilizer group defined in Eq.\eqref{eq:qdTableauRep.} and $T_q^A$ is the submatrix of $T_q$ taking only the entries supported on $A$. From the rank-nullity theorem we have 
\begin{equation}
\operatorname{rank}_q T_q^A + |\mathcal{G}(\mathcal{S}_{\overline{A}})| = N
\end{equation}
Now we have
\begin{equation}\label{eq:EEformula}
    S_A^\alpha = S_{\overline{A}}^\alpha = N - |A| - |\mathcal{G}(\mathcal S_{\overline{A}})| = \operatorname{rank}_q T_q^A - |A|.
\end{equation}
For a qudit graph state, the tableau representation takes the form $T_q = [I_N, T_G^w]$ as shown in Sec.~\ref{sec:quditStabilizer}. The truncated matrix $T_q^A$ is then
\begin{equation}
    T_q^A = \left(
    \begin{matrix}
    I_A & T^w_{AA}\\
    0_{\overline{A}} & T^w_{\overline{A}A}\\
    \end{matrix}\right).
\end{equation}
It is easy to see that $\operatorname{rank}_q T_q^A= \operatorname{rank}_q I_A + \operatorname{rank}_q T^w_{\overline{A} A}$, and $\operatorname{rank}_q I_A = |A|$. The entanglement entropy Eq.\eqref{eq:EEformula} can thus be written as 
\begin{equation*}
    S_A^\alpha = |A| + \operatorname{rank}_q T^w_{\overline{A} A} - |A| = \operatorname{rank}_q T^w_{\overline{A} A}.
\end{equation*}
This completes our derivation of Eq.\eqref{eq:quditEntropy} and  Eq.\eqref{eq:quditEEAdjMat}.

\section{Efficient graph state simulation algorithm}
\label{sec:algorithm}

In this appendix we introduce the algorithm we used to simulate the graph state on large size qudit system in this paper. Performing measurements after the complete construction of the size $L_x L_y$ circuit has complexity $\mathcal O(L_x^3 L_y^3)$ in the worst case. We use a dynamical evaluation method, which takes advantage of the fact that measurements only alters the stabilizers nearby. Hence we can construct the graph state up to size $L_y = 4$, and measure one layer in the middle and discard it from the memory. This does not affect the results and at the same time allows us to simulate the problem in a quasi one-dimensional setup, whose complexity is at most $\mathcal O(L_x^3 L_y)$. We thus can simulate system with large $L_y$.

Suppose we have a state with stabilized by the group $\langle s_1, s_2, \cdots, s_n \rangle$, with $n$ being the system size. Consider a measurement corresponds to stabilizer $s_o$. If all the stabilizers defining the state commute with the measurement, then the set of the stabilizers remain invariant. No operation is needed. Otherwise we order the pre-measurement state in the tableau representation  as the LHS of the following equation: 
\begin{equation}\label{eq:stabPreMea}
    T_q = \left(
    \begin{matrix}
    r(s_1) \\ 
    r(s_2) \\
    \vdots  \\
    r(s_k) \\
    r(s_{k+1})\\
    \vdots\\
    r(s_n)
    \end{matrix}
    \right) \xrightarrow[s_i' = s_i \quad i>k]{s_i' = s_i s_k^{\beta_i}\quad i<k}
    \left(
     \begin{matrix}
    r(s'_1) \\ 
    r(s'_2) \\
    \vdots \\
    r(s_k) \\
    r(s'_{k+1})\\
    \vdots\\
    r(s'_n)
    \end{matrix}\right)
\end{equation}
where $r(s_i)$ represents the row vector corresponds to stabilizer $s_i$. The first $k$ stabilizers are taken to be non-commutative with $s_o$. We perform a linear transformation to the RHS of Eq.~\ref{eq:stabPreMea}. For stabilizer $s_i$ with $i < k$, we multiple by $s_k^{\beta_i}$ to obtain $s_i'$ and require $[s_i', s_o] = 0$. Such $\beta_i$ always exist for the following reason. By the non-commutative condition, we define the phase $\alpha_i$ to be the phase: $s_i s_o = \omega^{\alpha_i} s_o s_i$. Then $s_i s_k^{\beta_i} s_o = \omega_q^{\alpha_i + \beta_i \alpha_k} s_o s_i s_k^{\beta_i}$. For $s'_i= s_i s_k^{\beta_i}$ and $s_o$ to commute, we have $\alpha_i + \beta_i \alpha_k = 0$ in $\mathbb{Z}_q$. Thus $\beta_i = q - \alpha_i ( \alpha_k^{-1} )$. (Note that $\alpha_k^{-1}$ always exists in $\mathbb{Z}_q$ when $q$ is prime.)
Now we have only one stabilizer -- $s_k$ -- that is non-commutative with $s_o$. After the measurement, $s_k$ is replaced by $s_o$. Therefore, we have the post-measurement tableau as in Eq.~\ref{eq:mea}
\begin{equation}\label{eq:mea}
T_q = \left(
     \begin{matrix}
    r(s'_1) \\ 
    r(s'_2) \\
    \vdots \\
    r(s_k) \\
    r(s'_{k+1})\\
    \vdots\\
    r(s'_n)
    \end{matrix}\right) \xrightarrow[]{s_k \mapsto s_o}
    T'_q = \left(
     \begin{matrix}
    r(s'_1) \\ 
    r(s'_2) \\
    \vdots \\
    r(s_o) \\
    r(s'_{k+1})\\
    \vdots\\
    r(s'_n)
    \end{matrix}\right).
\end{equation}

Let us analyze the computational complexity. In the worst case scenario, almost all the stabilizers do not commute with the measurement. It takes $\mathcal{O}(L^2_x L^2_y)$ (multiplication) operations to carry out the linear transformation in Eq.~\ref{eq:stabPreMea}. Conducting measurements in $\mathcal{O}(L_x L_y)$ bulk sites, the total number of operations is of order $\mathcal{O}(L_x^3 L_y^3)$.

In our work, we specialize to graph state, and our algorithm does not perform the measurement after the construction of the whole graph state. Instead, we perform measurements on the way of constructing the graph state in the $L_y$ direction. Specifically, we first construct the bonds (i.e. CZ gates operations) in Fig.~\ref{fig:cir} up to $L_y = 4$. The we conduct measurements on the second layer from the top and drop it from the computation. We proceed to construct the bonds on the fifth layer. We repeat this process until we reach the desired length of $L_y$. In the computation, the (dynamical) size of the circuit in the $y$ direction never exceeds $4$. The cost of measurement of each layer is thus $\mathcal{O}(L_x^3)$, according to the estimate above. The total cost is therefore $\mathcal{O}(L_x^3 L_y)$.

\section{The spin interactions in the Haar circuit}
\label{app:haar}

In this appendix, we provide more details about the spin mapping and the calculation of the spin interactions in Sec.~\ref{sec:haar}. 

We will not setup the random averaging facilities from scratch. Rather, we will first describe the formalism to compute the partition function in a 4-qudit example, and then repeat the calculation of the two body interaction terms in Ref.~\onlinecite{napp2020efficient} and finally work out the four-body interaction in our model. For more thorough study of the techniques and its broader applications, see for example Refs.~\onlinecite{Nahum_2017,zhou2019emergent,bao2020theory,jian2019measurementinduced,napp2020efficient,fan2020self}.

In the main text, we introduce two partition functions $\overline{Z_{\emptyset}}$ and $\overline{Z_{A}}$ in the definition of the quasi-entropy. These partition functions, after averaging over random gates, become a graph with permutation spins $(12)$ and $\I$ as vertices and the edges carry multiplicative weights. 

\begin{figure}[h]
\centering
\includegraphics[width=0.9\columnwidth]{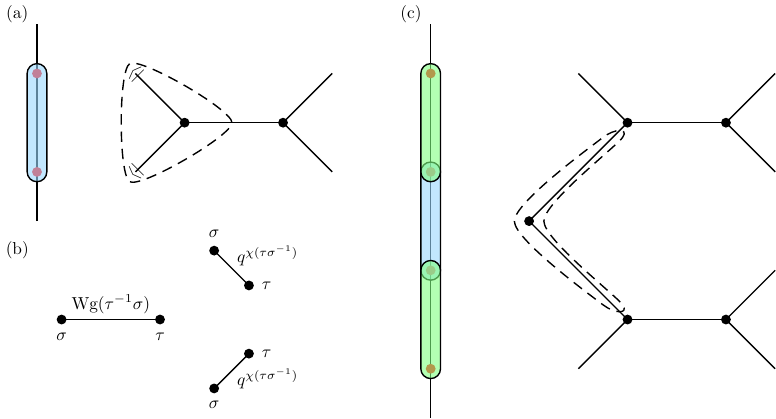}
\caption{General rule of random averaging and mapping to a spin model. (a) Averaging over a gate in the first layer. The gate is replaced by a four-leg tensor. The two 3-degree vertices are permutations spins. $\langle |$ represents contraction with the product initial state, which gives weight $1$. Summing over the spins in the dashed triangle contribute a weight $1$ to the quasi-entropy. (b) Rules of contracting spins. Horizontal bond carries a Weingarten function; bond with $45^\circ$ carries a weight of $q^{\chi(\sigma \tau^{-1})}$, where $\chi$ is the number of cycles in the permutation. (c) The random averaging after two layer. The blue region is the first layer. It gives a wedge of $90^\circ$ with a $\tau$ spin at the tip. The green region is the second layer. It gives a ``scattering'' diagram. We can integrate out the $\tau$ spin -- a 2-degree vertex to create a two-body interaction for the $\sigma$ spins of the second layer. }
\label{fig:first_two_layers}
\end{figure}

Let us consider a four-site example shown in Fig.~\ref{fig:first_two_layers}(c). In the first layer of evolution, there is only one unitary gate, which is applied to qudit 2 and 3. In the second layer of evolution, there are two unitaries applied to qudit 1 and 2, and qudit 3 and 4 respectively. The averaging of the each unitary produces a four-leg tensor shown in Fig.~\ref{fig:first_two_layers}(a). There are two 3-degree vertices, each hosting a spin denoted by Greek letter $\sigma$ or $\tau$. When we consider quantities involving $N$-copies unitaries and its complexity conjugation, these spins live in the $N$th permutation group. Since now only $\tilde{\rho}^2$ is involved, the spins can only take two values: $\I$ or swap $(12)$. The horizontal edge carries the Weingarten function\cite{weingarten1978asymptotic,gu_moments_2013}, which encodes the unitary invariance of the Haar ensemble. In our example, it has only two values, assuming the spins on the tips of the edge are $\sigma$ and $\tau$
\begin{equation}
\left\lbrace
\begin{aligned}
  & \text{Wg}(\sigma^{-1} \tau = \I) = \frac{1}{q^4 - 1} & \sigma = \tau\quad  \\
  & \text{Wg}(\sigma^{-1} \tau  = (12)) = -\frac{1}{q^2(q^4 - 1)} & \quad \sigma \ne \tau \\
\end{aligned} \right. . 
\end{equation}
The non-horizontal edge also contracts two spins. Its weight is $q^{\chi{\sigma^{-1} \tau }}$, the function $\chi$ counts the number of cycles in the permutation. In our case, it only has two values
\begin{equation}
\left\lbrace
\begin{aligned}
  & q^2 & \quad \sigma = \tau \\
  & q & \quad \sigma \ne \tau \\
\end{aligned} \right.. 
\end{equation}
The weights of the edges are summarized in Fig.~\ref{fig:first_two_layers}(b). 

\begin{figure}[h]
\centering
\includegraphics[width=0.9\columnwidth]{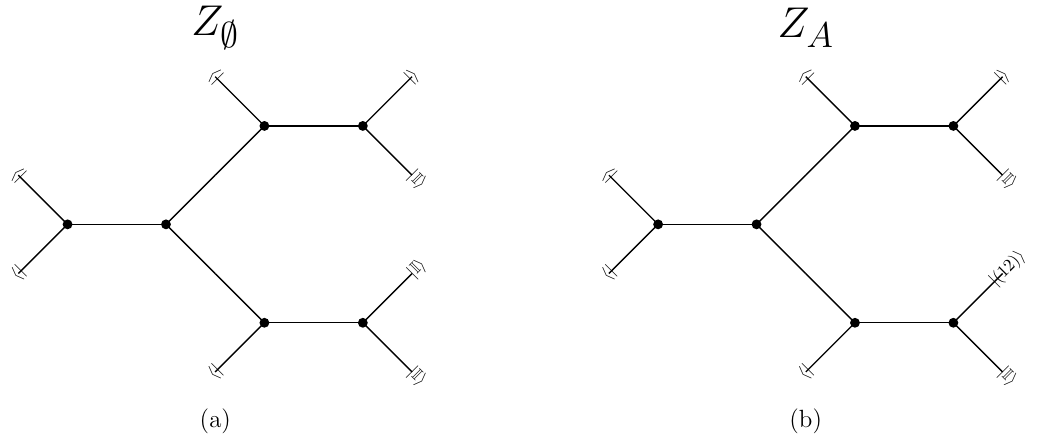}
\caption{Partition functions in a 4-qudit example. $\langle |$ and $| \rangle$ denotes the pure initial product state and finial projective measurement. The measurement is performed on the first site from the top and region A includes site 3 and 4.}
\label{fig:Z0_ZA}
\end{figure}
There are free legs in this graph (Fig.~\ref{fig:first_two_layers}(c)). Those on the left are contracted with the pure initial product state. Due to unitary invariance on the sites, the contraction with any product state evaluates to a constant value. We thus denote the state on each site as $\langle |$, see Fig.~\ref{fig:first_two_layers}(a). In fact the contraction of a spin with $\langle |$ gives $1$. The projective measurement on each site is similar, if we view the evolution backwards. We denote it as $|\rangle$. Contracting $|\rangle$ with the neighboring spinalso gives $1$. The remaining boundary conditions are determined by the partition functions. For $\overline{Z_{\emptyset}}$, all the remaining legs are contracted with a $\I$ spin. For $\overline{Z_{A}}$, the boundary spins are the same except that in unmeasured region complement to $A$, the leg is connected to a $(12)$ spin, see Figs.\ref{fig:Z0_ZA}(a)(b). 

We can see that compared to the uniform boundary condition in $\overline{Z_{\emptyset}}$, $\overline{Z_{A}}$ imposes a domain wall boundary state between $A$ and its complement, so that their ratio can be understood as a partition function of the domain wall. In this work, the boundary transition probed by the quasi-entroy, and ultimately the entanglement entropy of the boundary state, is a manifestation of the criticallity in the bulk. Therefore we will focus on the analysis of the spin interaction in the bulk. 

We follow Ref.~\onlinecite{napp2020efficient} to simplify the graph generated by the first two layers of unitaries. First of all, we can amputate the legs connecting to the $\tau$ spins in the first layer, see Fig.~\ref{fig:first_two_layers}(a). They evaluate to a constant, which appears in both $\overline{Z_{A}}$ and $\overline{Z_{\emptyset}}$ and cancels. Next, in the second layer, the remaining $\tau$ spin from the first layer is a 2-degree vertex. It can be integrated out to become an interaction between the $\sigma$ spins of second layer:
\begin{equation}
\left\lbrace
\begin{aligned}
  & q^4+q^2 & \quad \tau_1 = \tau_2 \\
  & 2q^3 & \quad \tau_1 \ne \tau_2 \\
\end{aligned} \right. 
\end{equation}
In terms of Boltzmann weight, it corresponds to an interaction
\begin{equation}
H = - \sum_{\langle ij \rangle }J_{\rm vert}  s_i s_j
\end{equation}
with $J_{\rm vert } = \frac{1}{2} \ln \frac{q^2 + 1}{2q}$.

\begin{figure}[h]
\centering
\includegraphics[width=0.9\columnwidth]{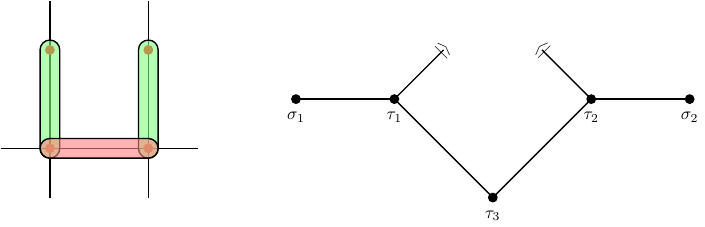}
\caption{Spin interactions when there are 3 layers of unitaries in total. Ref.~\onlinecite{napp2020efficient} adopted this construction of 3rd layers in a fractions of horizontal bond of the original lattice. After random averaging, unitaries in the third layer creates horizontal spin interactions between the $\sigma$ spins of the second layer.}
\label{fig:third_layer}
\end{figure}

Ref.~\onlinecite{napp2020efficient} continues to apply a third layer of unitaries among a fraction of horizontal bond of the original lattice (Fig.~\ref{fig:napp_gate}). The configuration of the two spins in a plaquette is shown in Fig.~\ref{fig:third_layer}. The third layer unitary is contracted with a projective measurement in the end. If we regard the (many-body) unitaries in the 3 layers as $U_1$, $U_2$, $U_3$, then the amplitude is
\begin{equation}
\langle \psi_{1} | U_3  U_2  U_1 |\psi_0\rangle 
\end{equation}
where $|\psi_0 \rangle $ represents the initial state, and $|\psi_1 \rangle $ represents the projected state in the end on meausured sites. Then this can be alternatively written as an overlap of $U_2 U_1 |\psi_0 \rangle $ and $U_3^\dagger | \psi_1 \rangle$. Averaging over the latter gives a $90^\circ$ 4-leg tensor connecting to the $\tau$ spins of the second layer. We consider those gates in the bulk and their free legs are measured. Again, we can amputate the measured legs and summing over the $\tau$ spins, see Fig.~\ref{fig:third_layer}. This gives us a weight about the $\sigma_1$ and $\sigma_2$
\begin{equation}
  \left\lbrace
  \begin{aligned}
    & \frac{1 + 2q + 4q^2 + 2q^3 + q^4}{q^2 (1+q)^2 (1+q^2)^2} & \quad \sigma_1 = \sigma_2 \\
    & \frac{2(1 + q + q^2)}{q (1+q)^2 (1+q^2)^2}  & \quad \sigma_1 \ne \sigma_2 \\
  \end{aligned} \right. 
\end{equation}
In terms of Boltzmann weight, it corresponds to an interaction
\begin{equation}
H = - \sum_{\langle ij \rangle }J_{\rm horiz}  s_i s_j
\end{equation}
with $J_{\rm horiz} = \frac{1}{2} \ln \frac{1 + 2q + 4q^2 + 2q^3 + q^4}{2q(1 + q + q^2)}$.

\begin{figure}[h]
\centering
\includegraphics[width=0.9\columnwidth]{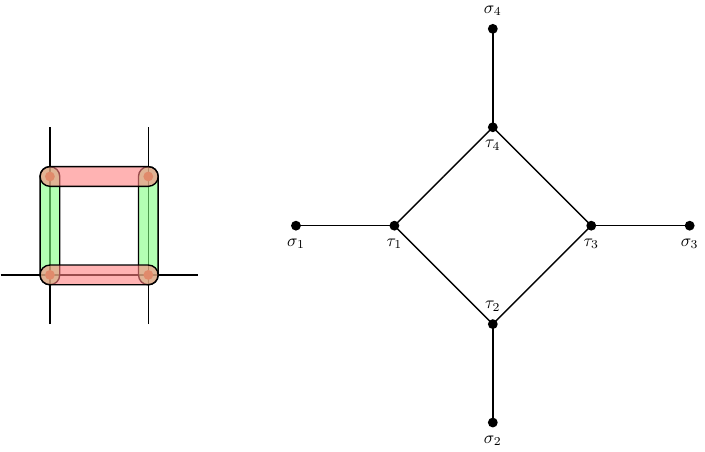}
\caption{Spin interactions when there are 4 layers of unitaries in total. In a plaquette, the first two layers create opposite spins $\sigma_1$ and $\sigma_3$. The third and fourth layers create opposite spins $\sigma_2$ and $\sigma_3$. They interact through contracting their corresponding $\tau$ spins.}
\label{fig:fig_four_body}
\end{figure}

Turning to our setup with four layers of unitaries, we call the (many-body) unitaries in these 4 layers as $U_1$, $U_2$, $U_3$ and $U_4$, then the amplitude after a full measurement is the overlap of $U_2 U_1 |\psi_0 \rangle $ and $U_3^\dagger U_4^\dagger | \psi_1 \rangle$. From this viewpoint, the averaging of the last two layers has the same structure as Fig.~\ref{fig:first_two_layers}(b) but this is stretched horizontally. We thus obtain spins on even horizon edges which interacts horizontally. When taking the inner product with the average of the first two layers, the tensor contraction in each plaquette is shown in Fig.~\ref{fig:fig_four_body}. The four body interaction has the expression
\begin{equation}
\text{Wg}( \sigma_1 \tau_1^{-1} ) J( \tau_1, \tau_3, \sigma_4 )J( \tau_1, \tau_3, \sigma_2 ) \text{Wg}( \sigma_3 \tau_3^{-1} ) 
\end{equation}
where $J( \tau_1, \tau_3, \sigma_4)$ is the 3-spin interaction after integrating out $\tau_4$. Integrating out the $\tau_1$ and $\tau_3$, the weight expression is given by a constant $\frac{(q^4 + 6q^2 + 1 )(q^2 - 1)^2}{8 q^4 (q^4 - 1) (1 + q^2)^2 }$ times 
\begin{equation}
\begin{aligned}
\frac{(q^2+1)^4}{(q^2 - 1)^2(q^4 + 6q^2 + 1 ) }s_1 s_2 s_3 s_4  \\
+ \frac{(q^2+1)^3}{(q^2 - 1)(q^4 + 6q^2 + 1 ) }(s_1 s_2 + s_2 s_3 + s_3 s_4 + s_4 s_1)\\
+ \frac{(q^2+1)^2}{(q^4 + 6q^2 + 1 ) }(s_1 s_3 + s_2 s_4)
\end{aligned}
\end{equation}

The constant will be canceled in the ratio of the partition function, so we drop it. Then we perform an large $q$ expansion about the four terms:
\begin{equation}
\begin{aligned}
1 + ( 1 + \frac{16}{q^4} ) s_1 s_2 s_3 s_4  \\
+ (1 - \frac{2}{q^2})(s_1 s_2 + s_2 s_3 + s_3 s_4 + s_4 s_1)\\
+ (1 - \frac{4}{q^2})(s_1 s_3 + s_2 s_4)
\end{aligned}.
\end{equation}
We can not find an exact match by a Boltzmann weight\footnote{A ansatz of the form $\prod_i (1+a_1 s_i s_{i+1}) \prod_i (1+a_2 s_i s_{i+2}) (1 + a_3 s_1 s_2 s_3 s_4 )/16$ gives the $q = \infty$ result at $a_1 = a_2 = a_3 = 1$. However expansions of the form $a_i = 1 - b_i / q$ produces the same expansion coefficients of $(s_1 s_2 + s_2 s_3 + s_3 s_4 + s_4 s_1)$ and $(s_1 s_3 + s_2 s_4)$ up to $\frac{1}{q^4}$.}. However an approximate Boltzmann weight 
\begin{equation}
\begin{aligned}
  \exp( &J_{1234} s_1 s_2 s_3 s_4 \\
  &+ J_{12}(s_1 s_2 + s_2 s_3 + s_3 s_4 + s_4 s_1)\\
  &+ J_{13} (s_1 s_3 + s_2 s_4) ) 
\end{aligned}
\end{equation}
with $J_{1234} \sim J_{12} \sim J_{13} \sim \ln q$ has a vanishing $\frac{1}{q}$ terms in the expansion. Hence in the large $q$ limit, we can view the four spins to effectively interact via ferromagnetic coupling $J_{1234}$, $J_{12}$ and $J_{13}$. 

When one of the gates in the first layer is absent, the corresponding 4-leg tensor in Fig.~\ref{fig:first_two_layers}(c) will not be present. Thus the interaction between the $\sigma$ spins will be absent. The same is also true for a vacant gate in the fourth layer. 

When there are vacant gates in the second or third layer, then one of the spins at the edge of the plaquette will be absent. If that spin was there, it would be interacting with two neighboring spins. Now the two neighboring spin will be interacting with the 3 other spins in the plaquette. In the large $q$ limit, the interaction will also be ferromagnetic. This interaction is more complicated than the ones considered in the main text. But we believe that the mechanism leading to the random bond Ising model remains the same.

\bibliography{ref}

\end{document}